\documentclass[natbib,smallextended]{svjour3}       
\smartqed  
\pdfoutput=1
\usepackage{graphicx}
\usepackage{color}
\usepackage{xcolor}
 \journalname{Space Science Reviews}
\newcommand{\aap}{{Astron. Astrophys.}}
\newcommand{\apj}{{Astrophys. J. Lett.}}
\newcommand{\apjl}{{Astrophys. J.}}
\newcommand{\prl}{{Phys. Rev. Lett.}}

\newcommand{\solphys}{{Solar Phys.}}
\newcommand{\mnras}{{Mon. Not. R. Astron. Soc.}}

\begin{document}
\title{Hemispheric Coupling:  Comparing Dynamo Simulations and Observations}
\author{A.A. Norton \and
P. Charbonneau \and
D. Passos}


\institute{A.A. Norton \at
              W.W. Hansen Experimental Physics Laboratory, Stanford University, CA, 94305, USA \\
              Tel.: 650-725-0474\\
              Fax: 650-725-2333\\
              \email{aanorton@stanford.edu}
\and
P. Charbonneau \at
        GRPS, D\'epartment de Physique, Universit\'e of Montr\'eal,
        C.P. 6128 Succ. Centre-ville, Montr\'eal, Quebec, Canada H3C-3J7\\
        \email{paulchar@astro.umontreal.ca }
\and
D. Passos \at
    CENTRA, Instituto Superior T\'ecnico, Universidade de Lisboa,
    Av. Rovisco Pais 1, 1049-001 Lisboa, Portugal\\
    GRPS, D\'epartment de Physique, Universit\'e of Montr\'eal,
        C.P. 6128 Succ. Centre-ville, Montr\'eal, Quebec, Canada H3C-3J7\\
    Departamento de F\'\i sica, Universidade do Algarve,
       Campus de Gambelas, 8005-139 Faro, Portugal\\
    \email{dariopassos@ist.utl.pt}}

\date{Received:  15 April 2014 / Accepted: 7 September 2014 }
\maketitle
\begin{abstract}
Numerical simulations that reproduce solar-like magnetic cycles can be used to generate long-term statistics.  The variations in north-south hemispheric solar cycle synchronicity and amplitude produced in simulations has not been widely compared to observations.  The observed limits on solar cycle amplitude and phase asymmetry show that hemispheric sunspot area production is no more than 20\% asymmetric for cycles 12-23 and that phase lags do not exceed 20\% (or two years) of the total cycle period, as determined from Royal Greenwich Observatory sunspot data.  Several independent studies have found a long-term trend in phase values as one hemisphere leads the other for, on average, four cycles.  Such persistence in phase is not indicative of a stochastic phenomenon.  We compare these observational findings to the magnetic cycle found in a numerical simulation of solar convection recently produced with the EULAG-MHD model. This long ``millennium simulation'' spans more than 1600 years and generated 40 regular, sunspot-like cycles.  While the simulated cycle length is too long ($\sim\,$40 yrs) and the toroidal bands remain at too high of latitudes ($>30^{\circ})$, some solar-like aspects of hemispheric asymmetry are reproduced. The model is successful at reproducing the synchrony of polarity inversions and onset of cycle as the simulated phase lags do not exceed 20\% of the cycle period. The simulated amplitude variations between the north and south hemispheres are larger than those observed in the Sun, some up to 40\%.  An interesting note is that the simulations also show that one hemisphere can persistently lead the other for several successive cycles, placing
an upper bound on the efficiency of transequatorial magnetic coupling mechanisms.  These
include magnetic diffusion, cross-equatorial mixing within latitudinally-elongated convective rolls (a.k.a. ``banana cells'') and transequatorial meridional flow cells. One or more of these processes may lead to magnetic flux cancellation whereby the oppositely directed fields come in close proximity and cancel each other across the magnetic equator late in the solar cycle.  We discuss the discrepancies between model and observations and the constraints they pose on possible mechanisms of hemispheric coupling.
\end{abstract}

\def\bfalpha{\setbox2=\hbox{$\alpha$}
\hbox{{$\alpha$}\hskip-.97\wd2
{$\alpha$}\hskip-.97\wd2
{$\alpha$}\hskip-.97\wd2 {$\alpha$}}}
\def\bfbeta{\setbox2=\hbox{$\beta$}
\hbox{{$\beta$}\hskip-.97\wd2
{$\beta$}\hskip-.97\wd2
{$\beta$}\hskip-.97\wd2 {$\beta$}}}
\def\bfgamma{\setbox2=\hbox{$\gamma$}
\hbox{{$\gamma$}\hskip-.97\wd2
{$\gamma$}\hskip-.97\wd2
{$\gamma$}\hskip-.97\wd2 {$\gamma$}}}
\def\bfomega{\setbox2=\hbox{$\omega$}
\hbox{{$\omega$}\hskip-.97\wd2
{$\omega$}\hskip-.97\wd2
{$\omega$}\hskip-.97\wd2 {$\omega$}}}
\def\bfOmega{\setbox2=\hbox{$\Omega$}
\hbox{{$\Omega$}\hskip-.97\wd2
{$\Omega$}\hskip-.97\wd2
{$\Omega$}\hskip-.97\wd2 {$\Omega$}}}
\def\bfxi{\setbox2=\hbox{$\xi$}
\hbox{{$\xi$}\hskip-.97\wd2
{$\xi$}\hskip-.97\wd2
{$\xi$}\hskip-.97\wd2 {$\xi$}}}
\def\bfvarepsilon{\setbox2=\hbox{$\varepsilon$}
\hbox{{$\varepsilon$}\hskip-.97\wd2
{$\varepsilon$}\hskip-.97\wd2
{$\varepsilon$}\hskip-.97\wd2 {$\varepsilon$}}}
\def\bftau{\setbox2=\hbox{$\tau$}
\hbox{{$\tau$}\hskip-.97\wd2
{$\tau$}\hskip-.97\wd2
{$\tau$}\hskip-.97\wd2 {$\tau$}}}

\def\derp#1#2{{\partial{#1}\over\partial{#2}}}
\def\dert#1#2{{{\rm d}{#1}\over{\rm d}{#2}}}
\def\ddt#1{{{\rm d}\over{\rm d}{#1}}}
\def\mpers{{\rm m}\,{\rm s}^{-1}}
\def\kgm3{{\rm kg}\,{\rm m}^{-3}}
\def\tadvec{\tau_u}
\def\tmagdif{\tau_\eta}
\def\emag{E_B}
\def\Pm{{\rm Pm}}
\def\Rm{{\rm Rm}}
\def\Re{{\rm Re}}
\def\Ro{{\rm Ro}}
\def\Co{{\rm Co}}
\def\St{{\rm St}}
\def\avB{\langle {\bf B}\rangle }
\def\avA{\langle {\bf A}\rangle }
\def\avJ{\langle {\bf J}\rangle }
\def\avU{\langle {\bf U}\rangle }
\def\av#1{\langle{#1}\rangle}
\def\turbA{{\bf a}^\prime}
\def\turbB{{\bf b}^\prime}
\def\turbU{{\bf u}^\prime}
\def\turbJ{{\bf j}^\prime}
\def\urms{u^\prime_{\rm rms}}
\def\emf{\bfxi}

\section{Introduction}
The magnetic solar cycle is thought to be caused by a dynamo mechanism operating in the solar interior, by which inductive fluid flows generate a strong reservoir of magnetism at the bottom of the convection zone and/or coherent magnetic structures in the bulk of this unstable region.
Many varieties of solar dynamo models have been designed based on varying levels of
dynamical and geometrical simplifications, and consensus has yet to arise as to which
best mimics solar behavior.   All have in common an assumed solar structure, differential rotation and magnetic diffusivity profiles, although there are significant differences in the diffusivity profiles used in the various models.  Some models invoke meridional circulation as highly important in the transport of magnetic fields,
while others consider magnetic diffusivity and/or turbulent pumping
to dominate over meridional circulation.

Alternately, dynamically
self consistent numerical simulations of dynamo action in rotating, stratified convective
turbulence can be run in order to compile statistics for dozens or more cycles, offering a unique laboratory to investigate how cycle outcomes depend on model assumptions.
Although such simulations currently cannot be run in a solar-like parameter regimes,
they can nonetheless reproduce some observed solar cycle behavior.

\begin{figure}
\includegraphics[width=1.00\textwidth]{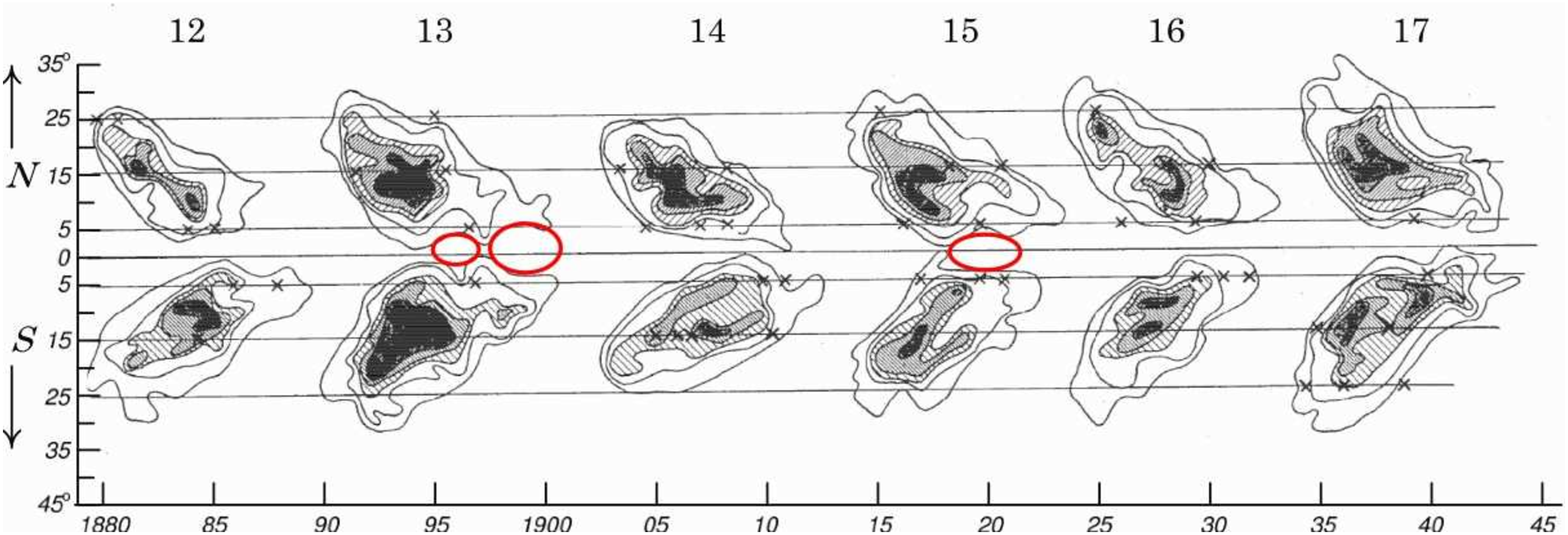}
\includegraphics[width=1.00\textwidth]{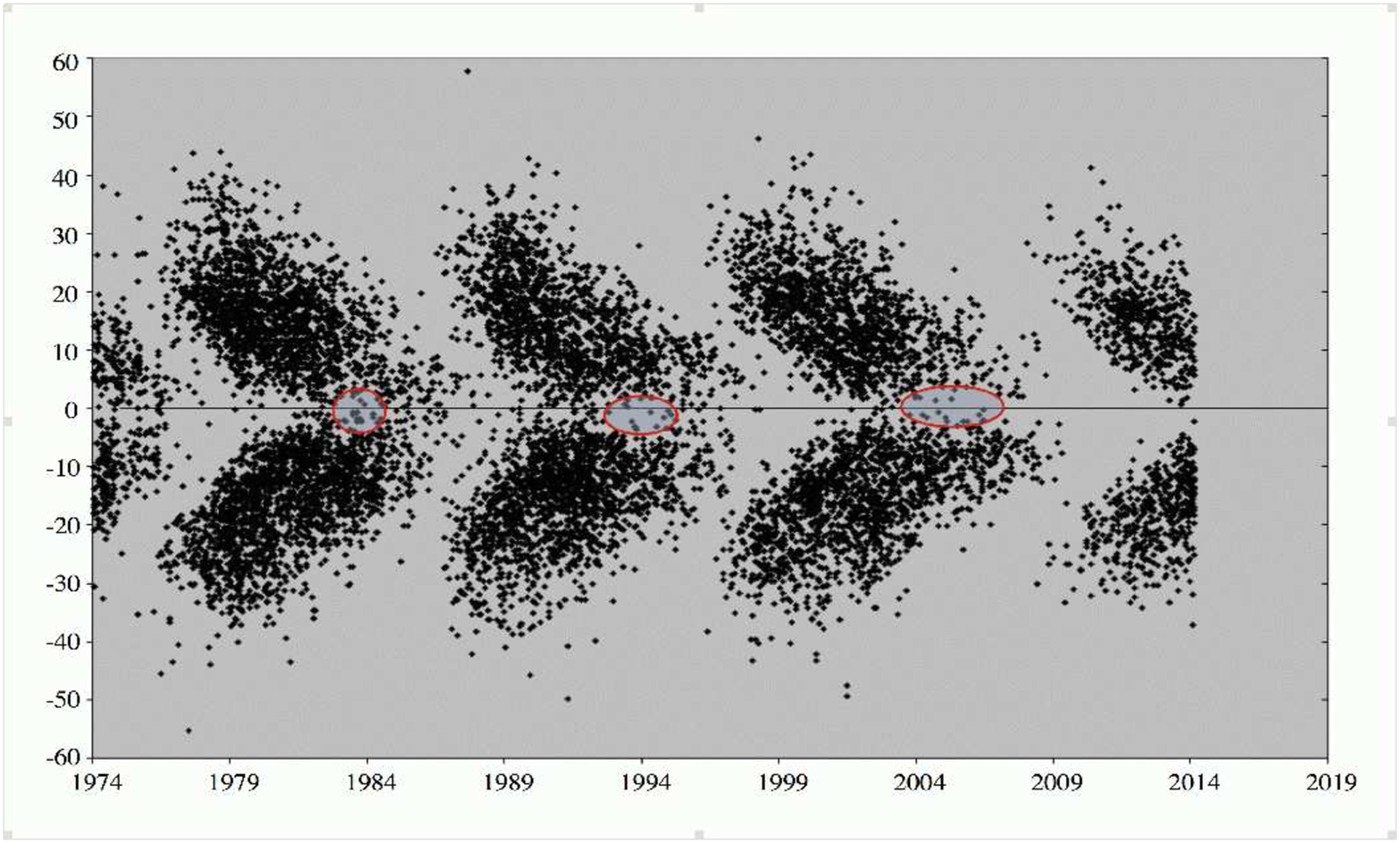}
\caption{Two versions of the butterfly diagram showing the latitudinal positions of sunspots as a function of time.  The top panel (Becker 1955) plots iso-contours of equal spot frequency for Cycles 12-17 from Greenwich Photoheliographic data while the lower panel plots each sunspot group position as an individual point from the end of Cycle 20 until February 2014. Data for the lower plot is the USAF/NOAA sunspot data also from the Greenwich observatory prepared by Jan Janssen (STCE: Solar-Terrestrial Centre of Excellence) at http://users.telenet.be/j.janssens/SC24web/SC24.html. Overplotted red ovals are drawn on the plot to highlight periods when less sunspots appear near the magnetic equator than the period just preceding the red ovals.  We propose that the time indicated by the red ovals may be when the leading edges of the N and S toroidal bands are undergoing a flux cancellation at depth. }
\label{fig:1}       
\end{figure}

Hemispheric synchronicity  and asymmetry is one aspect of cycle behaviour that can inform modelers about the relative importance of meridional circulation and other physical mechanisms that participate in hemispheric coupling.  In general, the Sun's hemispheres are considered moderately to strongly coupled since the wings of the butterfly diagram, see Fig 1., show a high degree of mirroring across the geometric equator.   Many models that reproduce the solar cycle assume a relatively
high magnetic Reynolds number, which means that magnetic diffusion plays a small part in the transport of the magnetic field, which is mainly achieved through meridional circulation. In this case, the N and S hemispheres can decouple. Then the dynamo and the subsequent sunspot cycle can progress independently in each hemisphere (Dikpati \& Gilman 2001; Chatterjee, Nandy \& Choudhuri 2004).

 Hemispheric asymmetry was noted at least as early as the works of Sp\"orer (1894) and Maunder (1904).  Waldmeier (1955) wrote ``the northern and southern hemispheres often show very unequal sunspot activity, and the inequality may persist for several years."  Newton and Milsom (1955)  also studied sunspot numbers and reported that the asymmetry in the activity of the two hemispheres from 1833 - 1954 did not appear to follow any definite laws.  In addition to a moderate amplitude asymmetry measured in sunspot production, and that sunspot maximum occurred at different times in the N and the S,  Babcock (1959) noticed that the polar fields reversed at different times: the south in mid-1957 and the north in late-1958.  Since then, significant hemispheric differences have been observed in many solar indices related to surface magnetism, and a more comprehensive list of these papers can be found in \S2.1.  In the past ten years, hemispheric differences have been observed in E-W zonal flows (Komm et al. 2014), N$-$S meridional flows (Komm et al. 2011) and Joy's law (McClintock \& Norton 2013).  The Sun's N$-$S asymmetry extends into the solar atmosphere, especially affecting the coronal magnetic equator and the deflection of the heliospheric current sheet, (see Virtanen \& Mursula (2014) and included references).  The southward shifted heliospheric current sheet has received attention as the ``bashful ballerina'' and has been persistently asymmetric with the northward component dominating the southward one during solar minimum for cycles 20-23 (Mursula \& Hiltula 2003) which is noteworthy.  One might not expect a hemisphere to be dominant instead of a magnetic polarity since the polarity in the hemispheres changed every 11 years.

In contrast with the observed hemispheric asymmetries detected over a half century ago in surface magnetism, only in the last two decades or so have dynamo modelling efforts included hemispheric differences.  Under the classical mean-field framework, the $\alpha$-effect -- the action of cyclonic convection on magnetic field lines -- depends only on the Coriolis force and stratification.
By argument of symmetry, the $\alpha$-effect should vanish at the equator allowing the possibility that the hemispheres become decoupled. The meridional circulation in flux transport dynamo models has, until recently, been thought to contain one cell per hemisphere (mirrored across the equator).  If there are multiple cells, they can evolve separately, and their contribution to the dynamo solution may translate into hemispheric asymmetries.  In the vast majority of these models only the magnetic diffusivity, $\eta$ is supposed to be isotropic  across all latitudes providing a reliable communication mechanism between both hemispheres.  Since Zhao et al. (2013) reported on the observation of multiple meridional cells with significant hemispheric flow differences, we must assume that the meridional flows do evolve separately to some extent in the different hemispheres.

Usually dynamo modelers look for mechanisms that can produce the asymmetric modulation that is observed in the solar cycle.  These mechanisms include stochastic forcing from convection (Hoyng et al. 1994), nonlinear effects caused by the magnetic backreaction of the induced magnetic field on the flows (Weiss 2010) or even other more controversial explanations such as the remnant of a  fossil or primordial field in the solar interior (Boyer \& Levy 1984).  Moreover, published modelling research addressing hemispheric asymmetries is somewhat scarce because, for simplification purposes, modelers tend to study dynamo solutions in only one hemisphere.  Early, keystone works of dynamo modellers to reproduce hemispheric asymmetry include the following.  Hoyng et al. (1994) focused on stochastic fluctuations in a 1.5D\footnote{In the technical jargon of this research field it is very common to use an integer number to represent spatial dimensions and 0.5 to make allusion to the time. Therefore 1.5D means one spatial dimension plus time.}, linear model and found phase and amplitude differences due to random fluctuations in the dynamo parameter $\alpha$ when applied with a correlation time on the order of the convective turnover time.  Ossendrijver et al. (1996) furthered this type of work proposing that the stochastic fluctuations could be a consequence of giant convective cells and found some interesting evidence of transequatorial activity (not unlike that being found in current day models with "banana cells").   Other works that concentrated on backreactions of magnetic fields into flow components to cause hemispheric asymmetries were Sokoloff \& Nesme-Ribes (1994)  and Tobias (1997) who studied the effects on differential rotation.  Another consideration is that an active coupling occurs deep in the interior when the leading edges of the N and S equatorward propagating toroidal bands ``meet up'' or come within a close enough range for flux cancellation and interaction to occur.

For the purpose of this paper, we review the literature dedicated to the observed hemispheric coupling (or lack thereof) in terms of amplitude and timing.  We then discuss which models have been used to reproduce the observed hemispheric coupling but focus mainly on recent results from EULAG-MHD modeling efforts.  We discuss which parameters or assumptions in the models most directly affect hemispheric coupling and address the following questions: are the observed hemispheric differences simply due to the possible stochastic nature of some dynamo ingredients?  In the light of recent modeling results, which coupling mechanisms contribute significantly to maintaining the symmetry of the butterfly diagram?


\begin{figure}
  \includegraphics[width=1.00\textwidth]{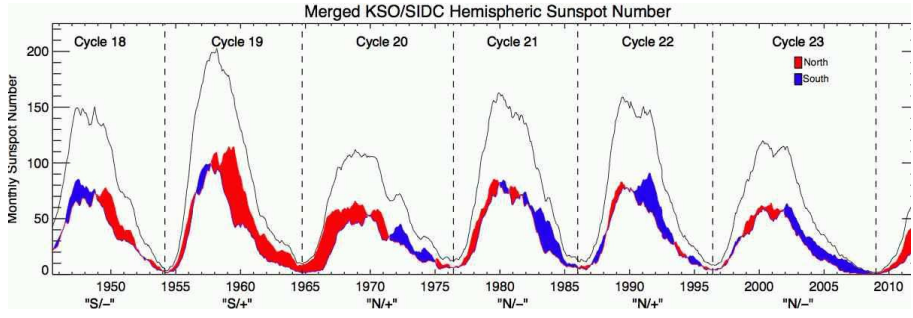}
\caption{International sunspot number by hemisphere combined from data provided by KSO and ROB/SIDC (e.g., Temmer et al. 2006) from 1945-2012. The N (red) and S (blue) hemispheric sunspot numbers are shown against the total (black). The shading of the difference between the N and S sunspot numbers indicates the excess between them. Dashed vertical lines are drawn at sunspot minima to delineate the cycles. The tags at the bottom indicate the hemisphere with most spots in the ascending phase of the cycle and the polarity of the magnetic field being advected into that hemisphere in the declining phase of the cycle. This figure reproduced from McIntosh et al. (2013).}
\label{fig:mcint1}       
\end{figure}

\begin{figure}
  \includegraphics[width=1.00\textwidth]{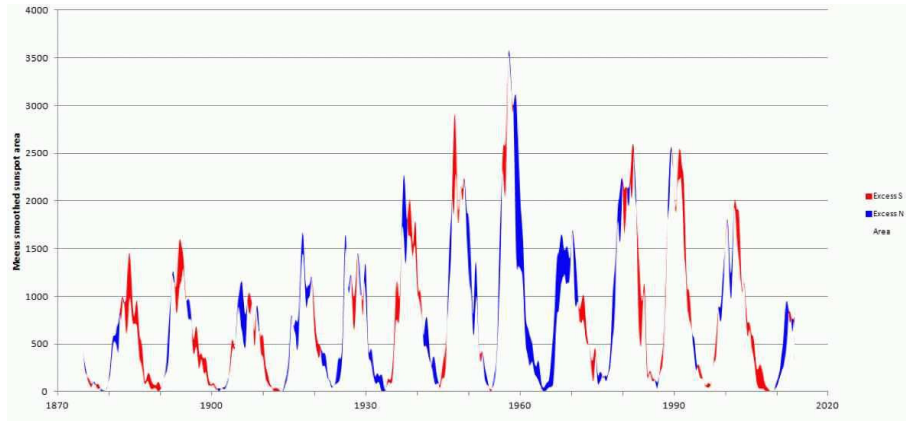}
\caption{Sunspot area by hemisphere combined from data provided by KSO and ROB/SIDC (e.g., Temmer et al. 2006) from 1876-2014. The excess N (blue) and S (red) (note: opposite colors compared to colors in Fig.1)  hemispheric sunspot areas are shown.  This figure produced by J. Janssens (STCE: Solar-Terrestrial Centre of Excellence).  }
\label{fig:Jans}       
\end{figure}

\section{Observations of N$-$S hemispheric asymmetry}
\subsection{Amplitude of Asymmetry Based on Sunspot Data}

Temmer et al. (2006) created a sunspot data set separated by hemispheres from the Kanzelh\"ohe Solar Observatory (KSO) and Skaltnat\'e Pleso Observatory for 1945-2004 (cycles 18-23).  The Solar Influences Data Center (SIDC, now the World Data Center-SIDC) has hemispheric sunspot numbers from the Royal Observatory Belgium (ROB) from 1992 to present.  These data have greatly assisted in the examinations of hemispheric asymmetry.  Temmer et al. (2006) report that the solar-cycle minima from 1945-2004 were in phase while the maxima, declining and increasing phases were clearly shifted. They also found that the N$-$S asymmetry, based on an absolute N$-$S asymmetry index, was enhanced at cycle maximum, which is in contradiction with results obtained by Joshi and Joshi (2004) who found the N$-$S asymmetry was greatest as solar cycle minimum but they were using a normalized index. The asymmetry index for a cycle in terms of any index A (such as sunspot area or numbers) is considered normalized, $\Delta_{Norm}$, if amplitudes $A^N,A^S$ in the northern and southern hemispheres are normalized by the total in that cycle.  As expected, they are considered absolute, $\Delta_{Abs}$, when not normalized by the size of the cycle in that index.
\begin{eqnarray}
\Delta_{Norm}={A^N-A^S\over A^N+A^S}, ~~~~ \Delta_{Abs}={A^N-A^S}
\end{eqnarray}
Both the normalized and absolute asymmetry indices can be problematic. The absolute asymmetry index provides strong amplitudes around the times of sunspot maxima while the normalized asymmetry index provides strong signals around the times of minima.

Several studies of the statistical significance of hemispheric asymmetry have been carried out. Carbonell et al. (1993) report that in most cases, the N$-$S asymmetry is highly significant and cannot be obtained from a distribution of sunspot areas generated randomly from binomial or uniform distributions.  Temmer et al. (2006) studied the monthly values of absolute asymmetry measures and found, for example, 63\% of the months during cycle 20 are asymmetric at the 99\% significance level. Carbonell et al. (2007) report that care should be taken when analyzing asymmetry due to different indices and binning techniques being employed.  Ballester et al (2005) remark that the time series generated from the normalized definition of asymmetry is misleading and that the correct asymmetry time series to be used is generated from an absolute difference between the northern and southern hemispheres.  For these reasons, the plots in this paper do not depict either the absolute or normalized index, instead we show the absolute asymmetry index by plotting a hemispheric excess in relation to the sunspot cycle, see Fig. 2 and 3. No periodicities in the time series of the asymmetry indices are determined other than inferences on the timescale of the solar cycle.
\begin{figure}
 \includegraphics[width=1.0\textwidth]{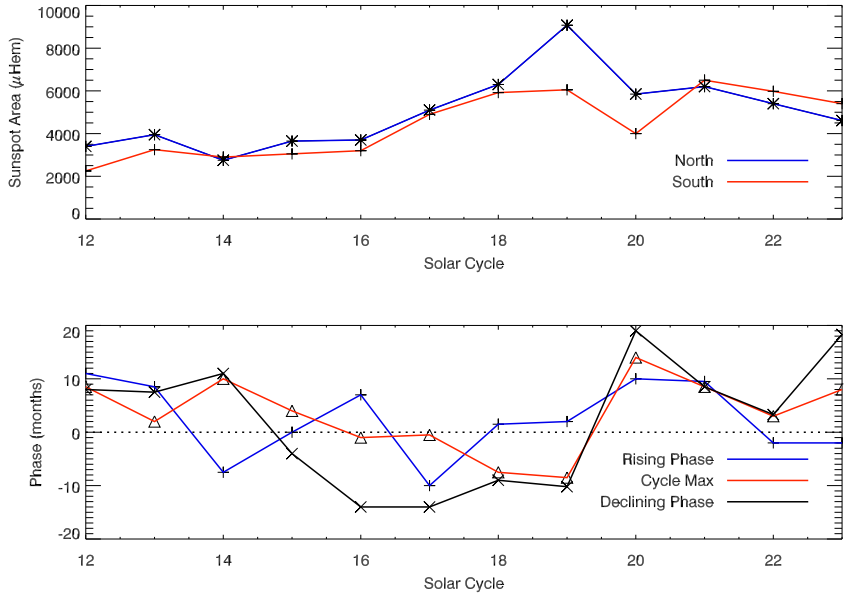}
\caption{Top:  The strength, or size, of sunspot cycles 12-23 shown as the summed sunspot area for the N (blue, asterisks) and S (red, crosses) hemispheres.  These data are from RGO sunspot area.  Bottom:  Hemispheric phase difference determined from RGO sunspot area for the rising, maximum and declining phases of each cycle.  Positive values indicate the N is leading the S.  Rising and declining phases are determined as the difference in time at which each hemisphere has produces 25\% of the maximum value for the rising and declining periods of the cycle.  Phases associated with cycle maxima are determined as differences in time each hemisphere takes to produce half the total sunspot area. }
\label{fig:NG1}       
\end{figure}
The asymmetry as measured by the hemispheric sunspot number is shown in Fig.~\ref{fig:mcint1} after smoothing with a 12-month window from 1945-2012 (McIntosh et al. 2013).  The authors report it as a hemispheric ``tug-of-war'' for dominance.  At any time, the differences in the hemispheric sunspot number is shown as red if the north has a higher sunspot number and blue if the south has a higher sunspot number.  At any give time, the magnitude of the shaded portion in Fig. 2 is the absolute hemispheric asymmetry index in sunspot area; red if the $\Delta_{Abs}$ in Equation 1 is positive or blue if negative.  A slightly different measure, the absolute asymmetry index determined from sunspot area, is shown in Fig.~\ref{fig:Jans} for a longer time span, 1870$-$2014. Unfortunately, the colors of N and S excess are reversed between Fig.~\ref{fig:mcint1} and \ref{fig:Jans} but the same behavior can be seen for cycles 18$-$23.  Due to historical sunspot counts being somewhat contentious, the absolute sunspot area is preferred over the sunspot number as a diagnostic for asymmetry.

It may be useful to report on the amplitude of hemispheric asymmetry exhibited over an entire cycle.  To do so, we report an average normalized asymmetry, $\Delta_{Norm}$, which is determined from the difference in sunspot areas produced by each hemisphere over an entire cycle, then normalized by the total sunspot area produced in that cycle.  For cycles 12-23, the average normalized asymmetry is 16\%, see Fig.~\ref{fig:NG1} (Norton \& Gallagher 2010).  Using sunspot number instead of sunspot area, the average normalized asymmetry for cycles 12-23 is slightly different, $\sim$11\% (Norton \& Gallagher 2010).  Cycle 19 and 20 are the outliers showing much more hemispheric asymmetry than any other cycles.  The N$-$S difference in sunspot area, $\sim$3000 $\mu$Hem, divided by the total sunspot area produced in cycle 19 by both hemispheres, $\sim$15000 $\mu$Hem, is 20\%.  Cycle 20 is similarly very asymmetric while half of the cycles show a much smaller, roughly 10\%, difference.  We provide an observed limit on the hemispheres not being more than 20\% asymmetric in sunspot area.

\subsection{Hemispheric Phase Lag Based on Sunspot Data}

Researchers calculate the hemispheric phase lag in a few different manners.  McIntosh et al. (2013) finds the phase lag (reported in units of months) by a cross-correlation of a 120-month window of the N and S sunspot area time series, see Fig. 5.  Norton \& Gallagher (2010) found that the cross correlation method of determining phase lag was not optimal because it took the entire solar cycle into consideration, whereas it is evident that the phasing of the hemispheres can change midway through the cycle.   They smoothed the sunspot area data over 24 months and normalized the resultant curves, then determined a hemispheric phase lag for the rising, maximum and declining epochs of each cycle.  The rising phase is defined as the difference in time at which each hemisphere reaches 25\% of the maximum hemispheric sunspot area for that cycle (see Fig. 1 in Norton \& Gallagher 2010).  Similarly, the declining phase is defined as the difference in time when the sunspot area has fallen to 25\% of the maximum for that cycle and hemisphere. For cycle maxima, Norton \& Gallagher (2010) summed the total sunspot area produced in each hemisphere and identified when half the total was produced. Hemispheric differences between these times are reported as the phases in units of months and plotted in Fig. 4, lower panel.   Zolotova et al. (2010) measure the phase differences of sunspot cycles in the two hemispheres using the Cross-Recurrence Plot (CRP) method (Marwan et al. 2007), where two time series are embedded in the same phase space.  They extracted phase information from sunspot data from the past four hundred years and show the persistence of phase over several cycles, see Fig. 6.

\begin{figure}
  \includegraphics[width=1.00\textwidth]{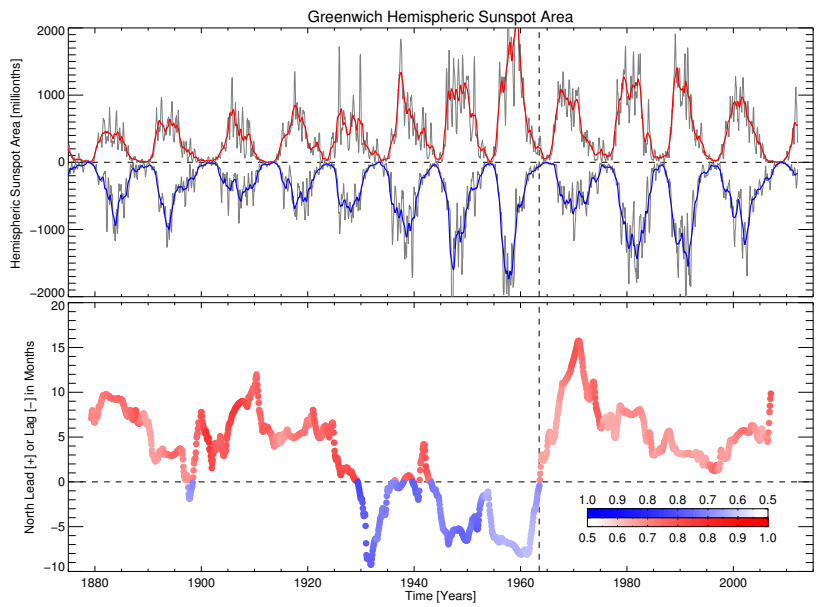}
\caption{Top and bottom: the Greenwich sunspot areas differentiated by hemisphere and the 120-month-averaged cross-correlation lag. Increasing color depth indicates the magnitude of the cross-correlation coefficient while the color itself dictates whether the north (red) or south (blue) is leading. This figure is reproduced from McIntosh et al. (2013), showing a proclivity towards the hemispheric lead to last for four cycles. }
\label{fig:mcint2}       
\end{figure}

\begin{figure}
  \includegraphics[width=0.95\textwidth]{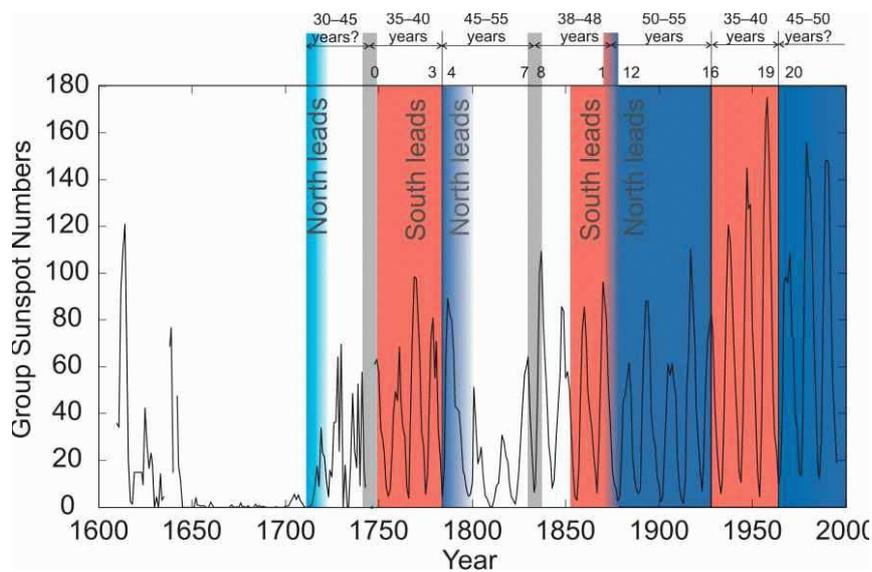}
\caption{Zolotova et al. (2010) plotted the yearly group sunspot numbers versus time.   Blue indicated periods in which the northern hemisphere precedes the southern one.  Red indicates that the southern hemisphere is leading in time. Vertical lines indicate a sign change of phase difference.  When it is uncertain, gray bars indicate the probable time of a sign change. }
\label{fig:zolo1}       
\end{figure}

Zolotova et al. (2010) and McIntosh et al. (2013) independently came to similar conclusions about the hemispheric phase differences of the sunspot cycle.  Namely:  \textbf{If the N$-$S phase difference exhibits a long-term tendency, a memory of sorts, it should not be regarded as a stochastic phenomenon.}  Mcintosh et al. (2013) found a persistent phase lag that lasted roughly four cycles. Specifically, the N led for cycles 12$-$15, the S led for cycles 16$-$19 and then the N again led for cycles 20$-$23, see Fig.~\ref{fig:mcint2}.  Zolotova et al. (2010) investigated the phase difference of the sunspot cycles in the two hemispheres for a much longer duration as shown in Fig.~\ref{fig:zolo1}, using historical records from Staudacher, Hamilton, Gimingham, Carrington, Sp\"orer, and Greenwich observers, as well as the sunspot activity during the Maunder minimum reconstructed by Ribes \& Nesme-Ribes.  Zolotova et al. (2010) show that during the last 300 years, the persistence of phase-leading in one of the hemispheres exhibits a secular variation. Changes from one hemisphere leading to the other were registered near 1928 and 1968 as well as two historical ones near 1783 and 1875, see Fig.~\ref{fig:zolo1}.

This is not the whole story, though.  Other researchers (Li et al. 2002; Norton \& Gallagher, 2010) parsed the solar cycle into distinct periods of minimum, rising, maximum, and declining times, then found the phase relationship between the hemispheres during these times. (Of course, the smaller the time period analyzed, the smaller the statistical sample of data becomes, especially during cycle minimum). Li et al. (2010) found that less than half of the minimum time periods during cycles 12$-$23 (partial data for cycle 23) were significantly out of phase.  In contrast, 75\% of the cycles were significantly out of phase at cycle maximum.  Norton \& Gallagher (2010) report the hemispheric phase difference for cycle 12$-$23 for the rising, maximum and declining cycle times, see Fig.~\ref{fig:NG1}.   The phases shown for the declining periods are consistent with the picture painted by McIntosh et al. (2013) and Zolotova et al. (2010), with the last 4 cycles showing the N leading and hemispheric switches occurring at the end of cycle 19, and the cycles 16-19 being led by the S, etc. However, there is not a persistent 4 cycle phase found in the rising periods of the cycles.   It is not clear from these reports whether the hemispheres truly are in\--phase more often at cycle minimum and become out\--of\--phase as the cycle progresses, or whether the lack of data and the shorter duration of rise time simply make the phase difficult to ascertain at minimum and during the beginning of a solar cycle.   It is worth keeping in mind, though, that the phase lag between the hemispheres may be set in the duration of each cycle, as only the sign of the phase is persistent for 4 cycles, not the phase value itself.  The phase lags found by Norton \& Gallagher (2010) for Cycles 12-23 vary from 0$-$19 months translating into an upper limit of $\sim$15\% of the cycle length. Zolotova et al. (2010), who studied historical sunspot cycles back to 1750, found that the phase lag did not exceed 2 years, so we place an upper limit on the phase lag to be 20\%.

Hemispheres out of phase with one another may manifest in different latitudinal averages of sunspot locations. Phase-leading was found to be anticorrelated with the latitudinal distribution of the sunspots by Zolotova et al. (2009, 2010).  They describe that the sunspots in the leading
hemisphere show a butterfly wing emerging at slightly lower latitudes as compared to the butterfly wing of the  delayed hemisphere, where sunspots emerged at
slightly higher latitudes. The asymmetric placement of the butterfly wings suggests a magnetic equator offset from the geometric equator (Pulkkinen et al. 1999) which may result from one hemisphere leading the other in sunspot activity.

\subsection{Polar Magnetic Fields}
Polar fields are equally important since they present hemispherically asymmetric lag times of the field reversal and the different polar flux amplitudes. However, polar field reversals out of phase are not necessarily the best indicators of the hemispheres being out of phase because the high scatter in the tilt angles of bipolar sunspot pairs
means that an emergence at high latitudes of an active region with a significant poloidal component (high tilt angle) and large amount of flux can influence the polar field reversal significantly.  Basically, the low amount of flux needed to reverse the poles means that the stochastic nature of flux emergence and tilt angles can affect it greatly (see Hathaway et al., this volume).
However, the surface dipole is used for predicting solar cycles and is a window into the possible magnetic behavior deep inside the convection zone.

\begin{figure}[ht!]
\includegraphics[width=1.0\textwidth]{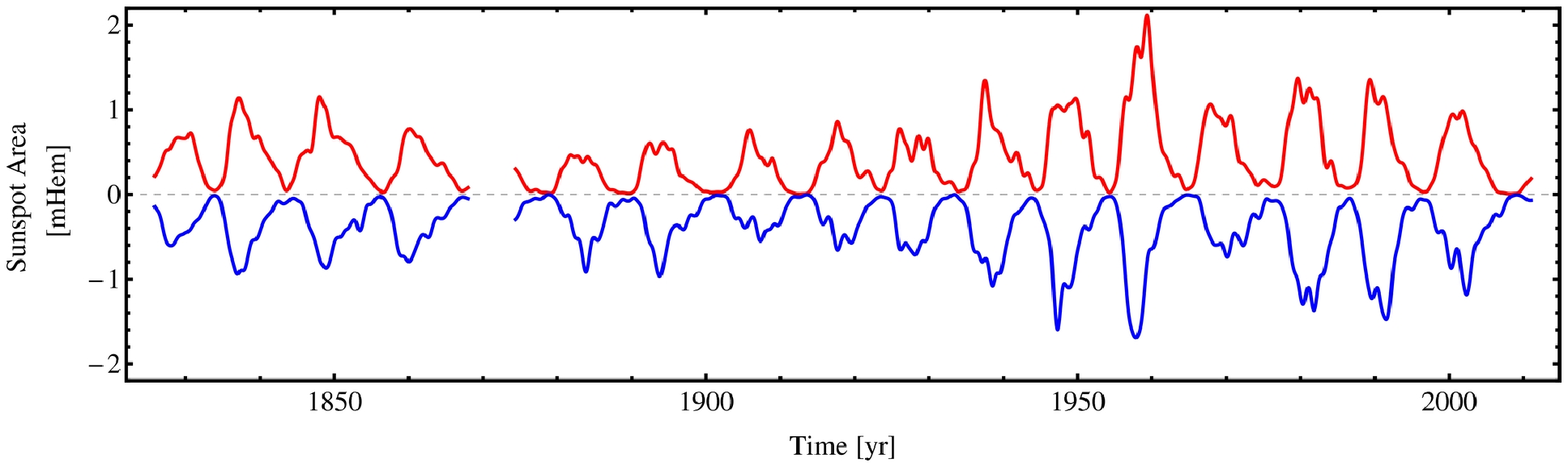}
\includegraphics[width=1.0\textwidth]{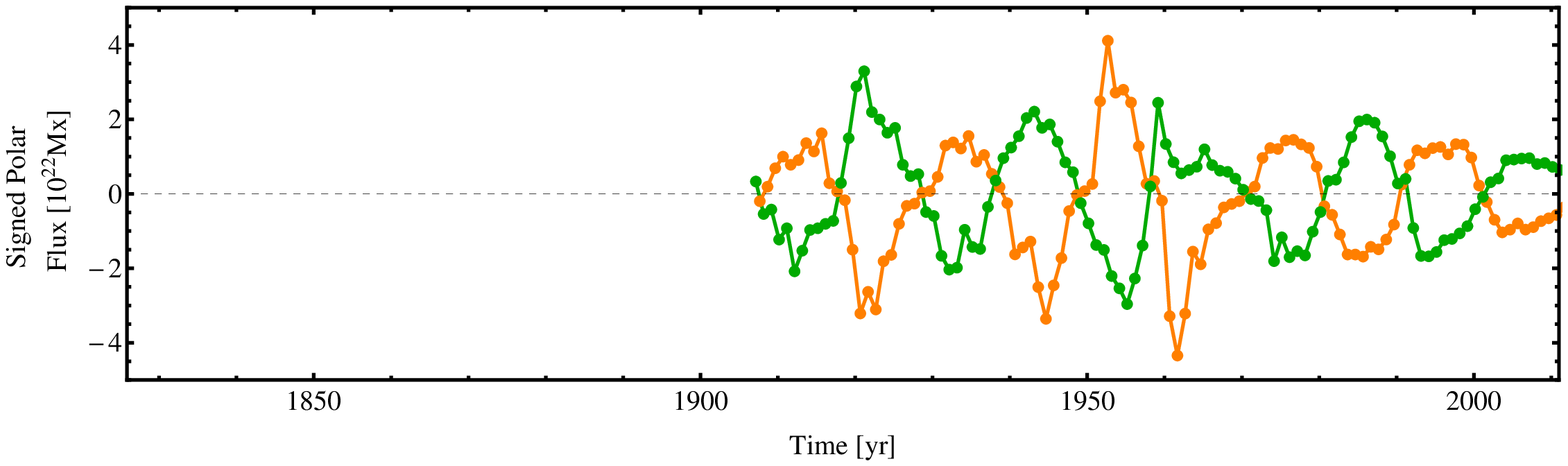}
\caption{Top:  Smoothed daily sunspot area for the northern (red line) and southern (blue line) hemispheres. The gap corresponds to missing data on cycle 11. Bottom: Polar flux (based on magnetic and polar faculae observations) for the northern (orange) and southern (green) hemispheres. Adapted from \cite{Munoz-Jaramillo2013b}. }
\label{fig:SSA+PF}       
\end{figure}

 Individual hemispheric data also show advantages over whole hemispheric data
for prediction purposes. For example, \cite{Munoz-Jaramillo2013a, Munoz-Jaramillo2013b},
showed that using the hemispheric polar flux as a precursor for the amplitude of the
following cycle yields better results than when using whole sun measurements.
\cite{Munoz-Jaramillo2013a} combine four campaigns observing the polar faculae from Mt. Wilson Observatory.  The data are combined and calibrated with MDI /SOHO to obtain a consistent polar flux database.  Using the polar faculae as a proxy for magnetic flux, \cite{Munoz-Jaramillo2013a} are able to provide a reliable time series of polar field dynamics from 1906 till 2013.   In Fig.~\ref{fig:SSA+PF}, we show the polar data used in these studies with an extended timeline of sunspot data also provided for reference.   The top panel of this figure shows smoothed daily sunspot areas for each hemisphere.  Cycles 7$-$10 were calibrated to RGO total area and SSN by Andr\'{e}s Mu\~{n}oz-Jaramillo using a compilation of Schwabe's data by Arlt (2013).  The details of the calibration and being prepared by Rainer Arlt and will be forthcoming later this year.   For cycles 12 through 23, the data was reduced by Andr\'{e}s Mu\~{n}oz-Jaramillo using the data of \cite{Balmaceda2009}. The polar flux curves are generated from the database of polar faculae observation observed at Mt. Wilson Observatory and calibrated from \cite{Munoz-Jaramillo2012}.  The details of this calibration are out of the scope of this review and we show the Fig~\ref{fig:SSA+PF} for illustration purposes only.

How asymmetric and out of phase can the poles become?   Babcock (1959) reported that the south polar field reversed its sign between March and July, 1957 and the north polar field, however, remained positive until November, 1958.  He states that for ``more than a year, the unexpected peculiarity was presented of two poles with the same sign''.  Cycle 19, then, had polar reversals that occurred eighteen months apart.  Durrant \& Wilson (2002) report that the northern polar field reversal occurred five months before the south in Cycle 23 from analyzing Kitt Peak and Mt. Wilson Observatory data.  Benevolenskaya (2007) confirmed this.  Using Wilcox Solar Observatory (WSO) data, collected since 1975, one can estimate the polar reversal times for Cycles 21-24 using field strength data smoothed over eighteen months (see http://www.solen.info/solar/polarfields/polar.html and Table \ref{tab:1}).  However, the times of polar reversal determined from the smoothed, low\--resolution WSO data do not match the times of polar reversal determined from the higher\--resolution Kitt Peak and Michelson Doppler Imager data.  One might expect the lags (differences in times of reversal) to vary using different data, but in the case of Cycle 23, the hemisphere that reversed first is not in agreement.  Determining polar field reversal times can be tricky since it depends upon the defined latitude above which the summed flux would be either negative or positive as well as any polar field interpolation methods employed (Sun et al. 2011).

\begin{table}
\caption{Polar reversal lag times in months.  Cycle 19 from Babcock (1959) while Cycles 21$-$23 are from Wilcox Solar Observatory. {$^{*}$Durrant and Wilson (2002) and Benevolenskaya (2007)
report that the north reversed first using Kitt Peak and MDI data.
There is a discrepancy since in the smoothed WSO data, it would appear the south reverses first.   }}
\label{tab:1}       
\begin{tabular}{lll}
\hline\noalign{\smallskip}
Cycle & Leading Hemisphere & Lag (months) \\
\noalign{\smallskip}\hline\noalign{\smallskip}
19 & South & 18 \\
21 & North & 5 \\
22 & North & 14 \\
23 & South (North)$^{*}$ & 5 (5) \\
24 & North & 13 \\
\noalign{\smallskip}\hline
\end{tabular}
\end{table}

\subsection{Miscellaneous: Phase Correlations with Cycle Characteristics, Dynamics}
Other research turned up null results:  Norton \& Gallagher (2010) report no correlation between hemispheric phase lag and duration of cycle while Murak\"ozy \& Ludm\'any (2012) find no relationship between phase lag and strength of solar cycle.   Norton \& Gallagher  (2010) also sought, but did not find, a correlation between the estimation of flux transport across the equator late in one cycle (based on Kitt Peak magnetogram data and an assumed surface diffusivity) and the phase of the hemispheres in the rising phase of next solar cycle.  In terms of dynamics, Chowdhury, Choudhary \& Gosain (2013) identified mid-duration and shorter periods in the asymmetry time-series data including a Rieger-type period ($\sim$155 days).  Antonucci, Hoeksema \& Scherrer (1990) used the low-resolution Wilcox Solar Observatory data to report that the central latitudes of sunspots were lower, the width of the latitudinal zone smaller and the rotation rate higher in the northern hemisphere  (15$^{\circ}$, 24$^{\circ}$ wide, 26.9 days synodic) compared to the southern hemisphere (26$^{\circ}$, 32$^{\circ}$ wide, 28.1 days) for cycle 21. A similar finding was reported using Mt. Wilson cycle 20 data.   The sunspots appearing at lower latitudes in the N for Cycle 20 and 21 corroborates Zolotova et al. (2010) findings that the leading hemisphere produces sunspots at a slightly lower latitude than the delayed hemisphere.

\subsection{Meridional and Zonal Flows}
Local helioseismology techniques, such as ring diagram (Hill 1989) and time-distance (Duvall et al. 1993) analysis, are able to determine non-symmetric latitudinal structure in the solar interior.  These analyses have been used to measure distinct hemispheric differences in the meridional flows and zonal flows at a given time and depth in the interior, see Komm et al. (2011) and Komm et al. (2014) and references included.  These measured flow asymmetries provide further quantitative constraints on the dynamo simulations in that the simulations must reproduce asymmetries within the range observed.

The extent of the hemispheric coupling as determined by helioseismology is as follows.  Zonal flows are seen as bands of faster and slower E-W flows that appear years prior to the appearance of activity on the solar surface.  Zonal flow patterns may be caused by enhanced cooling by magnetic fields.  Meridional flows are oft-considered the crucial ingredient which sets the rate at which the toroidal magnetic band (and sunspots) move equatorward.  Komm et al. (2011) determined both the zonal and meridional flows as a function of latitude for Cycle 23 (years 1996$-$2010), see Broomhall et al (2014), this volume.  They showed that meridional flow at 10$-$15 Mm in the northern hemisphere weakens in 2005 at 35$^{\circ}$ N latitude  just before the northern surface magnetic contour disappears in 2006 (see Fig 11, Broomhall et al., this volume).  Similarly, the southern hemisphere shows this behavior 2 years later in 2007 at 35$^{\circ}$ S latitude  just before the southern magnetic contour disappears in 2008.  This $\sim$2 year hemispheric phase lag observed in both the surface magnetism and the meridional flow is certainly suggestive.

Komm et al. (2014) also investigated the behavior of the zonal flows as a function of latitude for the time period of 2001$-$2013 from the surface to a depth of 16 Mm using GONG and HMI (as opposed to GONG and MDI data in the Komm et al. (2011) paper).  Many hemispheric differences are evident in the zonal flows.  For example, they find the poleward branch of the zonal flow (at $\sim$50$^{\circ}$ latitude) is 6 m s$^{-1}$ stronger in the S at a depth of 10$-$13 Mm during cycle 23. In addition, Zhao et al. (2013) detected multiple cells in each hemisphere in the meridional circulation using acoustic travel-time differences.  The double-celled profile shows an  equatorward flow extending from approximately 0.92$-$0.83 R$_{\odot}$ with a speed of about 10 m s$^{-1}$. The poleward flow covers depths of 0.92$-$1.0 R$_{\odot}$ as well as 0.75$-$0.83R$_{\odot}$.  These flows show significant hemispheric asymmetry in a range of latitudes.  The profile asymmetry could be due to a phase lag in the hemispheres.  It begs the question: does the meridional profile as a function of depth in the southern hemisphere in 2013 look like the profile did in the northern hemisphere two years earlier?

A perturbation in the meridional flow (presumably by turbulence since the meridional flow is a weak flow strongly driven by convection) in one hemisphere (but not in the other) could set a phase lag between the migration of activity belts, and hence, the sunspot production, that persists for years. Although it is unlikely that the hemisphere that leads would be the same one for 4 cycles in this case.  Actively searching for correlated hemispheric asymmetric signatures in flows at depth and magnetic field distributions on the surface may provide insight as to which ingredients of the dynamo set the length and amplitude of the sunspot cycle.

\section{Dynamo Models: N$-$S asymmetries and hemispheric coupling\label{sec:dynamo}}

Mean-field electrodynamics is well-covered in a number of textbooks and review articles (e.g.,
Moffatt 1978; Krause \& R\"adler 1980; Ossendrijver 2003;
Brandenburg \& Subramanian 2005; Charbonneau 2014).
What follows is consequently brief, and focuses primarily on aspects of the theory concerned with turbulent transport of the mean magnetic field.

In mean-field and mean-field-like models of the solar cycle, the diffusive decay and
transport of magnetic fields is almost always modeled as an isotropic Fickian (linear) diffusion, with the diffusion coefficient assuming large values, meant to reflect the enhanced dissipation by the small-scale turbulent flow pervading the convecting layers. Even within the framework of mean-field electrodynamics,
the latter's impact on large-scale magnetic fields is in fact more complex.
The mean-field dynamo equations derive from the assumption of scale separation,
whereby the magnetic field and flow are decomposed as sums or large-scale (mean) and small-scale
(turbulent) components, ${\bf B}=\avB+{\turbB}$,
${\bf U}=\avU+\turbU$, and the definition of an averaging operator such that $\av{\turbU}=0$ and $\av{\turbB}=0$. Under these conditions, introduction of scale separation into the MHD induction equation and averaging leads to the mean-field induction equation:
\begin{equation}
\label{eq:mfmhd}
\derp{\avB}{t}=\nabla\times (\avU\times \avB +\emf -\eta\nabla\times \avB)~,
\end{equation}
where $\eta$ is the magnetic diffusivity (inversely proportional to the electrical conductivity of the plasma) and  $\bfxi=\av{\turbU\times\turbB}$ is the mean turbulent electromotive force (emf), appearing now in addition to the usual motional emf $\avU\times\avB$ associated with
the large-scale flow and magnetic field. This turbulent emf can be developed as a Taylor
series in terms of the mean magnetic field:
\begin{equation}
\label{eq:alphabeta}
{\cal E}_i = \alpha_{ij}\av{B}_j + \beta_{ijk}\derp{\av{B}_j}{x_k}+ ...~,
\end{equation}
where the tensors $\bfalpha$, $\bfbeta$, etc, may depend on the properties of the flow, but not on $\avB$.
Retaining only the first two terms in this expansion and separating the symmetric and antisymmetric parts of the $\alpha$-tensor as $\alpha_{ij}=\alpha_{ij}^S -\varepsilon_{ijk}\gamma_k$ then yields:
\begin{equation}
\label{eq:alphabetagamma}
{\cal E}_i = \alpha_{ij}^S\avB_j + [\bfgamma\times\avB]_i+\beta_{ijk}\derp{\avB_
j}{x_k}~,
\end{equation}
with $\gamma_i=-(1/2)\varepsilon_{ijk}\alpha_{jk}$. Upon substitution into eq.~(\ref{eq:mfmhd}), $\bfgamma$ emerges as a vectorially additive contribution to the large-scale flow $\avU$. This pseudo-flow is known as turbulent pumping, and acts as a transport
agent for $\avB$. In the case of near-homogeneous, near-isotropic turbulence, the isotropic (diagonal) component of $\alpha_{ij}^S$ and $\beta_{ijk}$ reduce to
\begin{equation}
\label{eq:alpha}
\alpha=-{\tau_c\over 3}\av{\turbU\cdot\nabla\times\turbU}~,
\end{equation}
\begin{equation}
\label{eq:beta}
\beta={\tau_c\over 3}\av{(\turbU)^2}~,
\end{equation}
and turbulent pumping becomes proportional to the gradient of turbulent intensity:
\begin{equation}
\label{eq:gamma}
\bfgamma=-{\tau_c\over 3}\nabla \av{(\turbU)^2}~,
\end{equation}
where $\tau_c$ is the correlation time of the turbulent flow, usually assumed to be commensurate with the convective turnover time.
The mean-field induction equation (eq.~(\ref{eq:mfmhd})) then becomes:
\begin{equation}
\label{eq:mfmhdbis}
\derp{\avB}{t}=
\nabla\times ((\avU+\bfgamma)\times \avB +\alpha\avB
-(\eta+\beta)\nabla\times \avB)~.
\end{equation}
Note that the isotropic part of the $\bfalpha$ tensor describes an electromotive force aligned with the mean field $\avB$, and the $\bfbeta$-tensor appears as an additive contribution to the magnetic diffusivity $\eta$, with order-of-magnitude estimates based on eq.~(\ref{eq:beta})
indicating that $\beta\gg\eta$. This is the turbulent diffusion introduced in virtually all mean-field and mean-field-like models of the solar cycle. Note already that $\beta$ captures
only one specific contribution of the small-scale flow to the transport and dissipation of $\avB$.

In the modelling of the solar cycle, the large-scale magnetic field $\avB$ is considered axisymmetric, and the natural averaging operator becomes a zonal average (i.e. in the $\phi$ direction).
The large-scale flow $\avU$ includes (axisymmetric) contributions from differential rotation and meridional flows.
Both are ultimately driven by turbulent Reynolds stresses and latitudinal temperature differences (see \S 5.3 in review by Karak et al., this volume)
and so are symmetric about the equatorial plane on timescales longer than the convective turnover time. Turbulent pumping is directed primarily downwards throughout the bulk of the convection zone (Tobias et al.~2001), but in the physical regime where convection is significantly influenced by rotation, a latitudinal, equatorward component of turbulent pumping also materializes, reaching speeds comparable to the meridional flow
(see, e.g., Ossendrijver et al.~2002; Racine et al.~2011; also Fig. 12 in Karak et al. review, this volume). This latitudinal component of $\bfgamma$ may contribute significantly to the observed equatorward drift of the sunspot butterfly diagram.

\subsection{Mean-field dynamo models\label{ssec:mfemods}}

In classical mean-field dynamo models of the solar cycle, differential rotation is producing the toroidal large-scale magnetic field, while the turbulent emf is responsible for the regeneration of the poloidal component. However, in the solar dynamo context, additional magnetic source terms may appear on the RHS of eq.~(\ref{eq:mfmhdbis}). In particular, the surface decay of bipolar active regions and subsequent poleward transport of decay products by surface diffusion and meridional flow can contribute to the dipole moment, thus acting as a source term for the poloidal field, a process known as the Babcock-Leighton mechanism. Various MHD instabilities of the strongly toroidal flux system accumulating in the tachocline, immediately beneath the base of the convection zone, can also act as sources of poloidal field. For further discussion
of these dynamo models see \S 4 in Charbonneau (2010).

With the large-scale flows and tensors $\bfalpha$ and $\bfbeta$ considered given, eq.~(\ref{eq:mfmhdbis}) becomes linear in $\avB$, and accepts eigensolutions with temporal dependence $\propto \exp(\,(\sigma+i\omega) t)$. The growth rates $\sigma$ for the lowest odd (dipolar) and even (quadrupolar) eigenmodes are often similar, with high diffusivity typically favoring the very lowest eigenmode (usually dipolar), while dominant transport by a quadrupolar meridional flow will often favor the quadrupolar eigenmode. The tendency for coexistence of dipolar and quadrupolar solutions can persist in the nonlinear regime, and in itself lead to hemispheric asymmetries and attendant modulation of the magnetic cycle's amplitude. The key is then to ensure that the non-dominant eigenmodes are continuously excited to some significant amplitude. It turns out that this can be achieved though a variety of mechanisms, reviewed in what follows.

\subsection{Stochastic fluctuations}

Since the dynamo resides in the solar convection zone, a very turbulent place, it is only natural to assume that the physical mechanisms that operate in this region can be perturbed by convective turbulence (Hoyng 1988).
The most common recipe applied to simulate this perturbative effect is to add stochastic fluctuations to the chosen physical mechanism, at a correlation time associated with the phenomenon's characteristic time-scale (Choudhuri 1992; Moss et al.~1992).
In mean-field dynamo models the usual targets are the $\alpha$-effect and the meridional circulation. By definition, the large scale meridional circulation arises from longitudinal averaging of the global velocity field.
The flow speed associated with this weak flow (of the order of a few m s$^{-1}$)
is much lower than the velocities of turbulent eddies, which makes it susceptible to fluctuations. The origin of fluctuations in the mean-field $\alpha$-effect is based on the same principle and are supported by some highly turbulent global 3D MHD simulations of solar convection
\citep{Ossendrijver2001, Racine2011}.
In addition, significant fluctuations in the Babcock-Leighton mechanism are also expected because of the large variation in the tilt angle distribution of bipolar solar active regions \citep{Longcope1996, DasiEspuig2010}. This type of modelling methodology can produce hemispheric asymmetric solutions because the fluctuations applied are usually spatially uncorrelated  (different fluctuations at different latitudes and depths or different in both hemispheres).

In \cite{Hoyng1994} the authors use a simplified 1.5D ($\theta, t$) mean-field dynamo and study the effects of random fluctuations in their $\alpha$ source term (mean helicity). These fluctuations are a function of latitude and are applied at a correlation time of the order of the turnover time of convection (much shorter then the solar cycle period). They show that by increasing the level of fluctuations they produce larger hemispheric asymmetries between the wings of their simulated butterfly diagram.
In this model the interaction between fundamental and higher order excited eigenmodes is the source of the hemispheric asymmetries observed. \cite{Ossendrijver1996} used an $\alpha \omega$ mean field dynamo in ($r, \theta, t$) to further test this idea. They propose that the origin of the stochastic fluctuations in the $\alpha$-effect could be a consequence of giant convective cells. Under certain parameter regime their model also produces transequatorial magnetic activity.

A different approach to the implementation of stochastic fluctuations was taken
by \cite{Goel2009}. These authors suggest that the random nature of the Babcock$-$Leighton
mechanism can make the poloidal field in one hemisphere stronger than the other inducing
 therefore an hemispheric asymmetry. They test this idea by feeding into a BL flux
 transport dynamo model observational polar flux data and find a good correlation
 between the simulated cycle and observation. Their methodology is analog to
 that presented in \cite{Dikpati2006} and sets a preliminary framework for using
 mean field models for prediction purposes.

\begin{figure}[h!]
 \center
 \includegraphics[width=\textwidth]{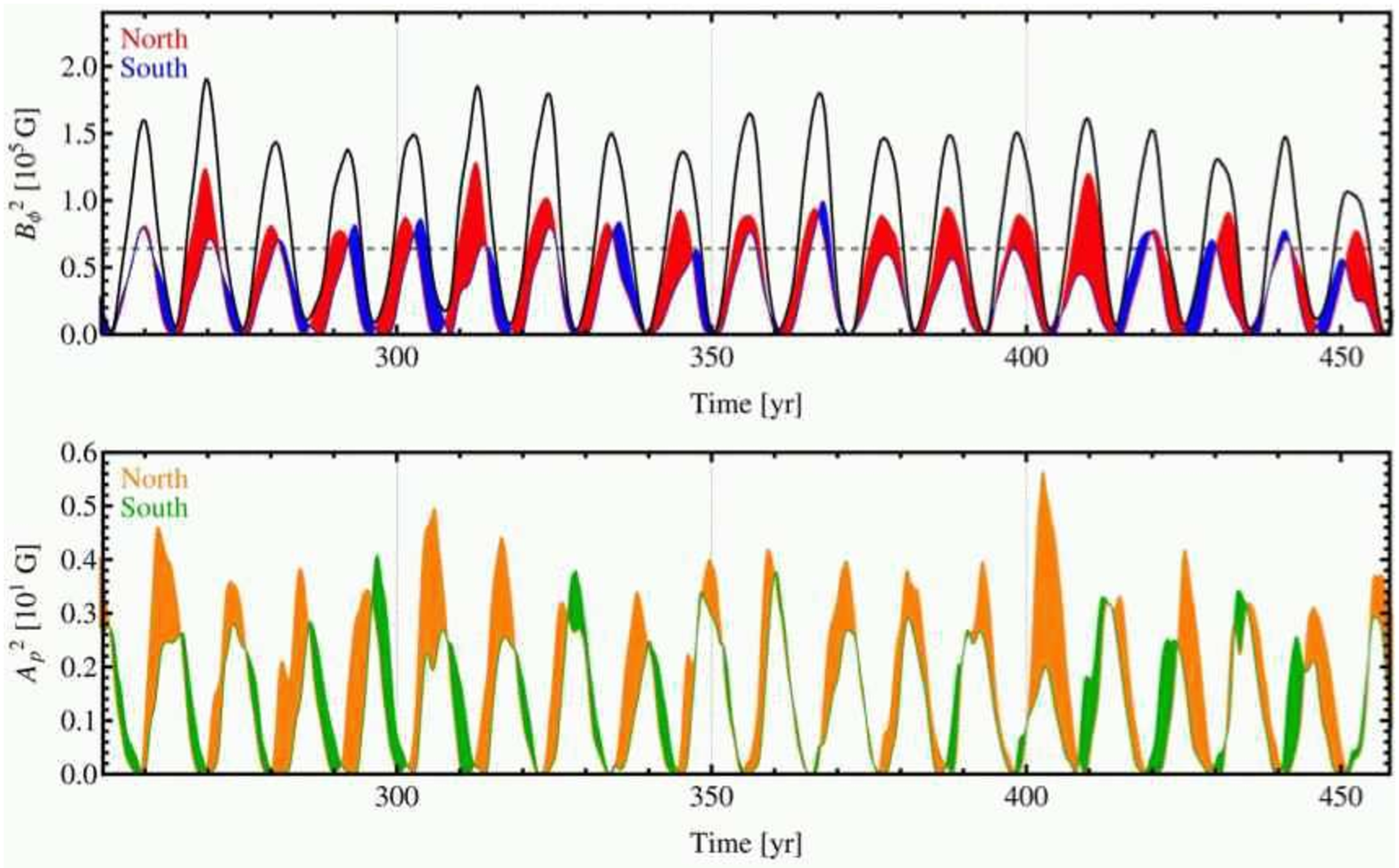}
 \includegraphics[width=1.08\textwidth]{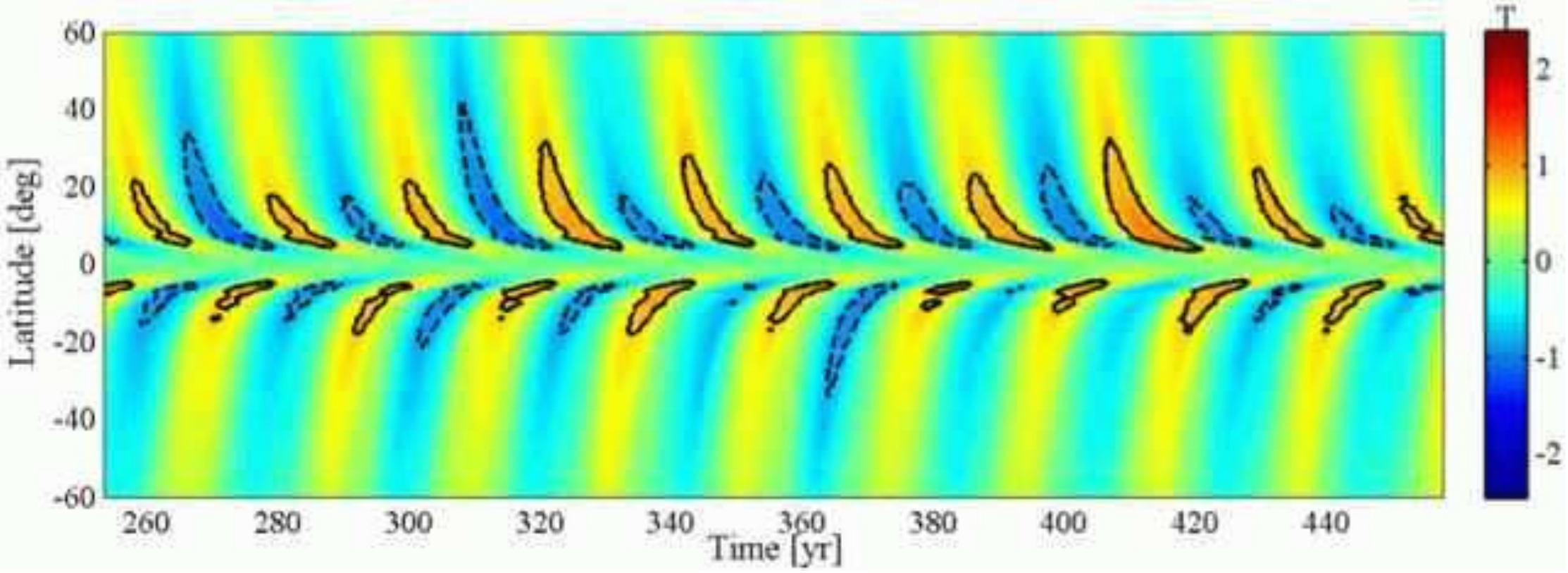}
 \caption{Sample of a simulation using the model of \cite{Passos2014a}. Represented here are
 in the top panel, the squared toroidal field measured just above the tachocline at
 $r=0.706\,R_\odot$ at active latitudes (14$^\circ$N (red) and 14$^\circ$S (blue)). The black thick line
 represents the sum of the two hemispheric contributions.
 The thin dashed line
 indicates the buoyancy threshold at $B_\phi=0.8\times10^5$ G.
 The middle panel shows the squared poloidal field measured at near surface layers at
 $r=0.96\,R_\odot$ and near the north (orange) and south (green) poles.
 The bottom panels displays
  the toroidal field measured at $r=0.706\,R_\odot$ as a function of latitude and time.
  The black contour lines enclose the regions where the buoyant emergence of toroidal field
  takes place (analog to sunspot emergence locations). Solid (dashed) contours represent
  positive (negative) polarity. In this sample solution, dipolar parity is maintained.}
 \label{fig:DP1}
 \end{figure}

More recently \cite{Passos2014a} developed a mean-field flux transport dynamo model that incorporates a dual poloidal source formalism, namely a Babcock-Leighton mechanism, $\alpha_{BL}$ that acts on strong toroidal fields that buoyantly rise to the near surface layers and a mean-field classical poloidal source, $\alpha_{MF}$ that acts only in weak toroidal fields that diffuse through the convection zone. The Babcock-Leighton mechanism was held as the leading source, i.e. $\alpha_{BL} > \alpha_{MF}$. Nevertheless the relative contribution from both terms is a bit more difficult to evaluate because while $\alpha_{BL}$ is confined to a thin region near
the surface, the $\alpha_{MF}$ although lower in amplitude spreads across a larger area, the bulk of the convection zone. The authors study several scenarios where stochastic fluctuations are added to these two source terms individually or simultaneously. According to flux tube simulations, e.g. Caligari et al. (1995), the rise time of these objects through the CZ is of the order of months. Since that is the time that flux tubes are exposed to turbulent buffeting, the authors chose a correlation time of 6 months for the fluctuations in the BL source term and 1 yr for fluctuations in the mean-field source term. As previously mentioned in the beginning of this section, the amplitude of the fluctuations are motivated by the large variation of the tilt angle distribution in active regions and by the eddy velocities distributions and large scale flows amplitude variations measured in several 3D MHD global simulations of solar convection.
This model returns a wide range of solutions, from high to low hemispheric coupling all the way to grand minima. 
For illustrative purposes we present in figure \ref{fig:DP1} an example of a solution computed with this model using the following level of fluctuations: 100\% for $\alpha_{MF}=0.4$ and 25\% for $\alpha_{BL}=21.0$ (see Passos et al. 2014 for details).

This dynamo model also incorporates a buoyancy algorithm that takes toroidal field exceeding
a buoyancy threshold at the base of the convection zone and places it in the near surface layers, at the same latitude. This is done to emulate the fast rise of flux tubes \citep{Chatterjee2004}.
It is then possible to plot the analog of a butterfly diagram with the
emergence latitudes highlighted (see figure \ref{fig:DP1}C). In this example we can observe that
the two hemispheres are moderately coupled but we also see phase lags between both hemispheres
(top panel) and different hemispheric emergence rates as seen in the different shapes of the butterfly wings of each cycle in figure \ref{fig:DP1}.

In this work the decoupling between hemispheres increases with the fluctuation level. The authors also find evidence that hemispheric decoupling is naturally associated with, and symptomatic of grand minima episodes, a result that also was proposed by \cite{Olemskoy2013}.
The next step for this model its a complete mapping of the parameter space coupled with
statistical constraints from long reconstructions of solar activity.

A more general scenario that considers a superposition of nonlinear quenching and
stochastic forcing, was presented by \cite{Mininni2002}. They subjected a 1.5D $\alpha \omega$
model, containing a nonlinearity in the form of $\alpha$-quenching to stochastic fluctuations
in this parameter. The solutions obtained with this model are in good agreement with observations and, according to the authors can be explained as the interaction of the deterministic part of the solution with overtones excited by the forced stochastic fluctuations. The asymmetries in their model arise as the combination of stochastic effects and deterministic coupling, and evolution of mixed-parity excited modes.

\subsection{Non-kinematic mean-field models\label{ssec:mfenonlin}}

Non-kinematic dynamo models that include some aspects of
the backreaction of the magnetic field on the large-scale flow components, can also show hemispheric asymmetries in their solutions. That is the case of the work by \cite{Sokoloff1994}, where the authors propose that the observed north-south asymmetry in the solar cycle amplitude and phasing could be a consequence of mix-parity solutions caused by nonuniform rotation.
In these mix-parity solutions, the dipole and quadrupole components can interact leading to hemispheric asymmetries. Also using a nonlinear $\alpha \Omega$ mean-field model (although more complex) that included the Malkus-Proctor effect (the feedback of the magnetic field into the differential rotation), \cite{Tobias1997} arrived at similar conclusions. Moreover, this author argues that relevant asymmetries only arise from mixed-parity solutions when the dynamo exhibits weak-field solutions (like in a grand minimum) and that the Malkus-Proctor effect has a more profound impact in the asymmetry for dipolar strong-field solutions. Building on the results and model of \cite{Tobias1997}, \cite{Bushby2003} studied solutions with higher degree of asymmetry, and discussed these in the more general context of solar-like stellar dynamos.

\cite{Pipin1999}, considered a weakly nonlinear $\alpha \Lambda$  mean field in a spherical shell. In this type of models the differential rotation is maintained by non-dissipative sources in
the angular momentum transport, the so-called $\Lambda$ terms, parameterizing Reynolds stresses.
Solutions produced by this model indicate that magnetic feedback on angular momentum fluxes produces a long term cyclic modulation that resembles the sunspot's Gleissberg cycle. The asymmetries between hemispheric activity comes from parity breaking, which in this model is always connected with breaking of the symmetry of differential rotation.

\subsection{Global MHD simulations of convection\label{ssec:eulag}}

The production of large-scale axisymmetric magnetic fields undergoing polarity reversals in strongly turbulent global (full-sphere) MHD simulations of solar convection is a recent development.
Computationally, the problem is quite challenging, as this large-scale magnetic field evolves on spatial and temporal scales much larger/longer than convection itself. When such large-scale fields build up and undergo polarity reversals, most often the latter occur at a irregular cadence, and often strong hemispheric asynchrony and asymmetries are present. Working with the
PENCIL code in spherical wedge geometry
K\"apyl\"a et al.~2012 find in some parameter regimes
a fairly regular cycle with a well-defined
period with sustained phase lags between hemisphere (see their Fig.~3);
but more typically the large-scale magnetic cycles materializing in their
simulations
are extremely irregular (see Figs.~4 and 5 in K\"apyl\"a et al.~2013).
The ASH simulations of Brown et al.~(2011) have produced short-period (a few yr)
cycles showing
large hemispheric asymmetry and asynchrony (see their Fig.~1B), a situation
improved upon to a significant extent in later
ASH simulations operating in a more strongly turbulent regimes (see Figs.~3, 6
and 15 in Nelson et al.~2013).
The recent global MHD simulations of Fan \& Fang (2014), on the other hand,
produce polarity reversals on decadal timescales and
well-synchronized across hemisphere, however with large cycle-to-cycle variations
in the half-period (see their Fig.~5).

To the best of our knowledge, the
global simulations of Ghizaru et al.~(2010) and Racine et al.~(2011) are currently
producing the most regular cycles, with polarity reversals well-synchronized across hemispheres.
Some tantalizingly solar-like secondary dynamical features are also reproduced, including rotational torsional oscillations and modulation of convective energy transport, both with the observed phasing and amplitude inferred on the sun (Beaudoin et al.~2012; Cossette et al.~2013). As a computational avatar of the real solar cycle, significant discrepancies remain: the cycle period is four times too long as compared to the sun, the deep-seated magnetic field is concentrated at mid-latitudes and exhibits very little equatorward
propagation, and the dipole moment is over ten times stronger than the solar one and oscillates in phase with the deep toroidal component, in contrast to the $\pi/2$ phase lag observed in the sun.

In what follows we focus on one specific EULAG-MHD simulation spanning 1600$\,$yr,
in the course of which 39 polarity reversals have taken place.
This is a low-resolution simulation ($128\times 64\times 48$ in longitude$\times$
latitude$\times$radius) spanning $0.604\leq r/R\leq 0.96$, mildly superadiabatic
above $r/R=0.711$ and strongly subadiabatic below, with stress-free impenetrable
upper and lower boundaries on which the horizontal magnetic field components are
forced to vanish.
The presence of a stably stratified fluid layer underlying the convectively unstable layers appears to be an important agent favoring
self-organization of the magnetic field on large spatial scales in these simulations (on this point see also Browning et al.~2006).
The magnetic cycle developing in this ``millennium simulation''
is analyzed in detail in Passos \& Charbonneau (2014), to whom we refer the interested reader for a full description of cycle characteristics, period-amplitude relationships, etc.

\begin{figure}
\begin{center}
\includegraphics[width=0.48\textwidth]{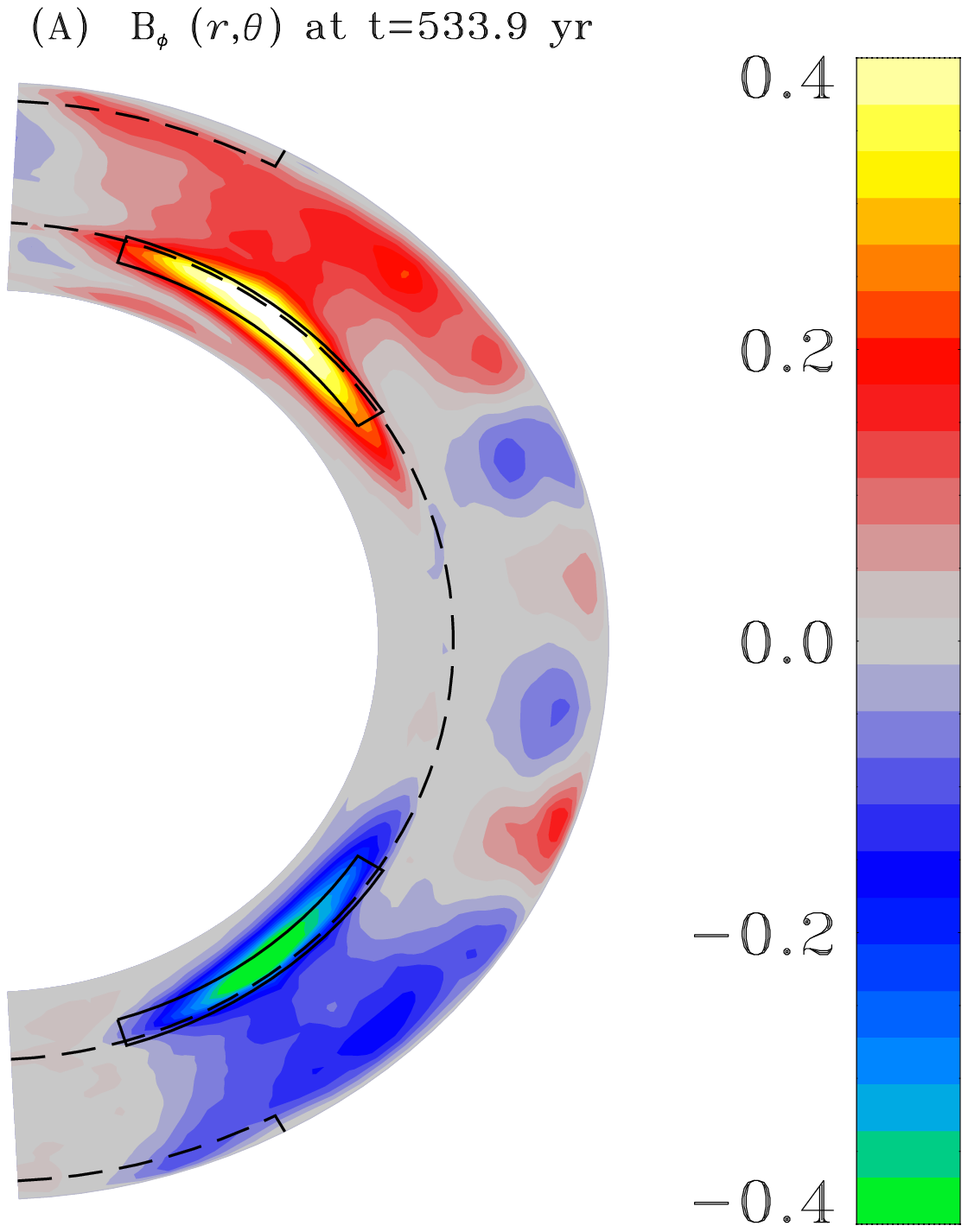}
\includegraphics[width=0.48\textwidth]{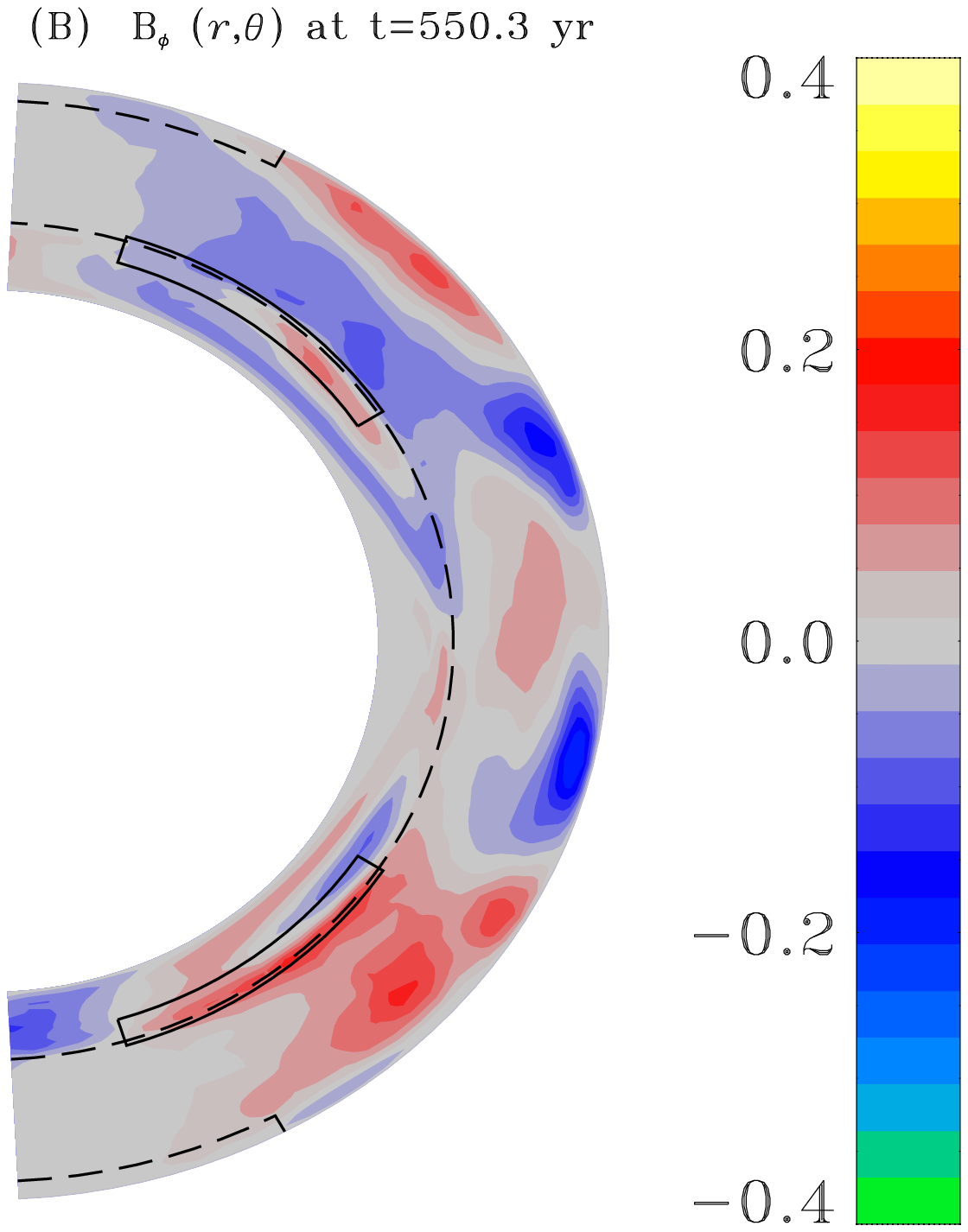}
\includegraphics[width=\textwidth]{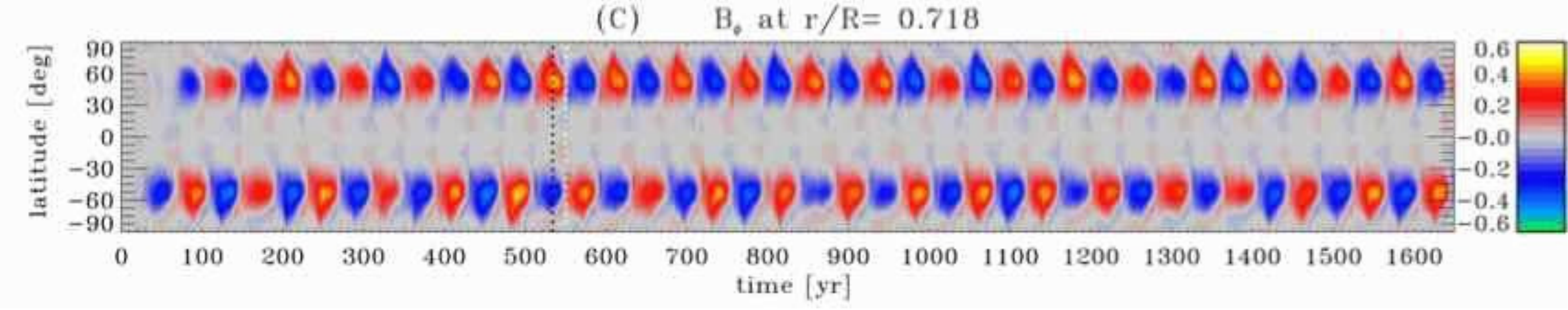}
\includegraphics[width=\textwidth]{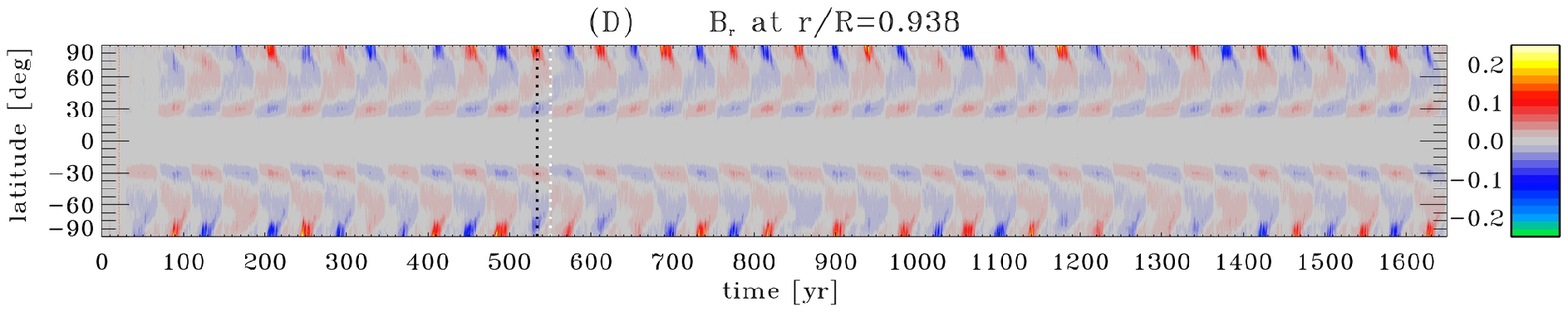}
\end{center}
\caption{Meridional snapshots of the zonally-averaged toroidal magnetic component taken at the time of solar maximum (A) and minimum (B). Panels (C) and (D) show time latitude diagrams for the toroidal component sampled at tachocline depth and the radial magnetic component in the near surface layers, respectively. The color scales codes the magnetic field strength, in Tesla, and the dashed circular arc in (A) and (B) indicates the boundary between the convectively unstable region and the underlying stable fluid layer, located at $r/R=0.718$;
the time-latitude diagram in (C) is extracted at that same depth, and that in (D) at $r/R=0.94$. The upper boundary of the simulation domain is at $r/R=0.96$. The meridional snapshots in (A) and (B) are extracted at the times marked in the time latitude diagrams with the dotted vertical lines, black for maximum and white for minimum. The boxes in these panels represent the areas of integration for the toroidal (solid box) and poloidal (dashed box) proxies. These results are taken from the global MHD simulation of
solar convection analyzed in Passos \& Charbonneau (2014). }
\label{fig:Dario0}
\end{figure}

Figure \ref{fig:Dario0}A,B shows two meridional cuts of the zonally-averaged toroidal magnetic field, at epochs of (A) cycle maximum and (B) minimum. The large-scale, axisymmetric toroidal magnetic component is antisymmetric about the equatorial plane, and accumulates immediately beneath the base of the convecting layers in response to downwards turbulent pumping, reaching there strengths of a few tenths of Tesla. Figure \ref{fig:Dario0}C shows a time-latitude diagram of the zonally-averaged toroidal magnetic field, extracted at the base of the simulated convection zone (dashed circular arc on Fig.~\ref{fig:Dario0}A and B). This is the simulation's equivalent of the sunspot butterfly diagram, under the assumption that the sunspot-forming toroidal magnetic flux ropes form in regions of highest magnetic intensity and rise radially to the surface. Figure \ref{fig:Dario0}D is a similar time-latitude diagram, this time for the evolution of the zonally-averaged radial magnetic field in the subsurface layers of the simulation ($r/R=0.94$, with the upper boundary of the simulation domain at $r/R=0.96$). The long term spatiotemporal stability of the magnetic cycle is well evidenced on these time-latitude diagrams.

\begin{figure}
\begin{center}
\includegraphics[width=\textwidth]{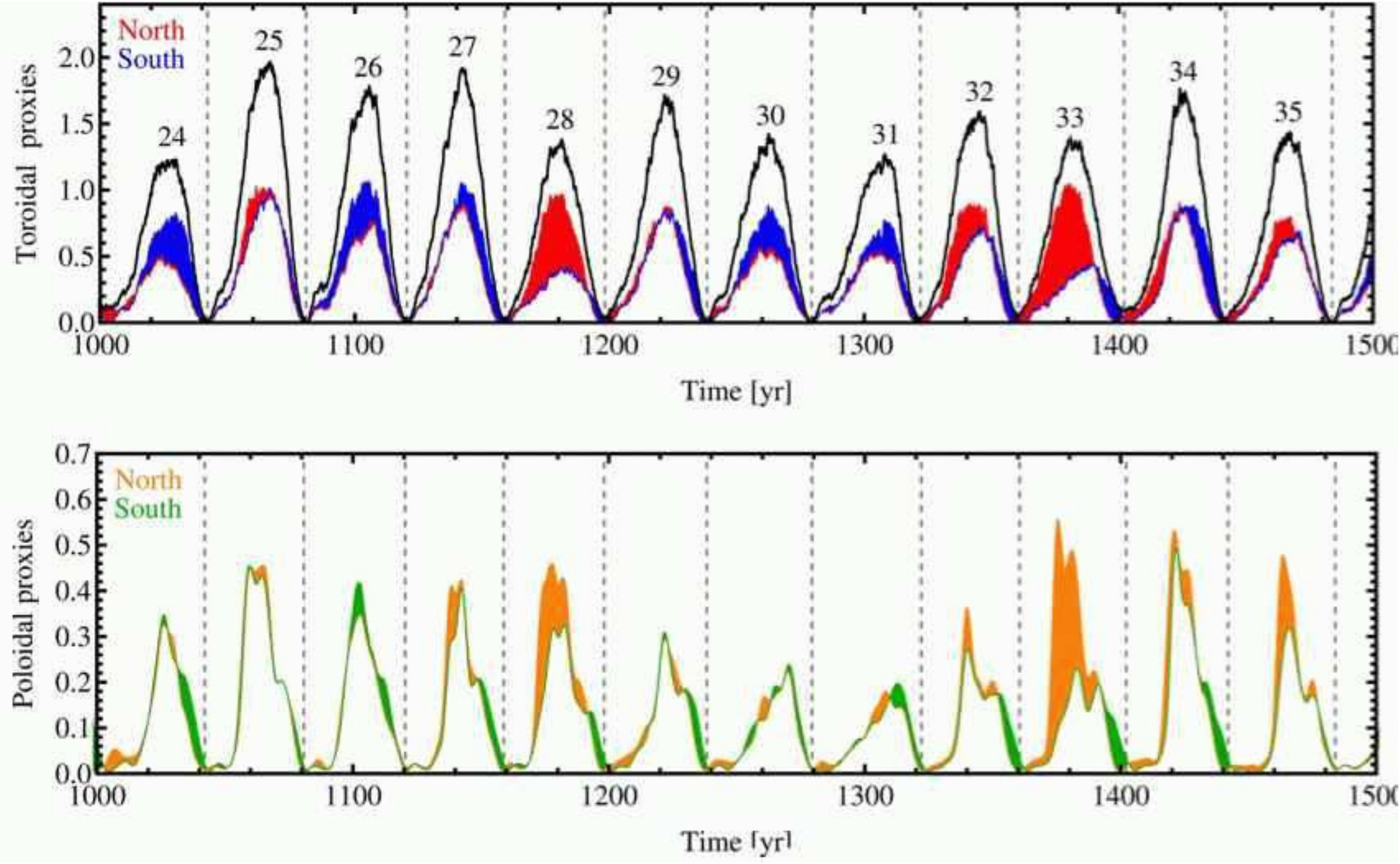}
\end{center}
\caption{A 500$\,$yr time series segments of the hemispheric toroidal (top) and poloidal (bottom) proxies in the northern and southern hemispheres,
in the EULAG-MHD ``millennium simulation'' analyzed by Passos \& Charbonneau (2014). The simulated cycles numbers are shown for reference.
Cycles are defined as spanning from one minimum to the next, and each corresponds
to one half of the underlying full magnetic cycle. The dashed vertical lines mark the time of minimum measured for the whole sun toroidal proxy (black thick line) that is the sum of both hemispheric contributions.}
\label{fig:Dario1}
\end{figure}

The sunspot number proxy constructed from the simulation output is the squared magnetic flux in an integration domain located at mid-latitudes and straddling the base of the convection zone, as indicated on Fig.~\ref{fig:Dario0} by the black box.
Similarly, a polar field proxy is constructed by integrating the radial magnetic field at high latitudes in the subsurface layers, indicated here by the dashed box (see also Passos \& Charbonneau 2014). Figure \ref{fig:Dario1} shows time series of these two proxies, calculated independently for the northern (red and orange) and southern (blue and green) hemispheres. Only a 500$\,$yr segment of the 1600$\,$yr simulation is plotted, covering twelve cycles. We follow the usual convention of numbering cycles from one minimum to the next, and refer to the corresponding time span as the ``cycle period'', even though the period of the underlying magnetic cycle covers two such cycles.

\begin{figure}
\begin{center}
\includegraphics[width=1.0\textwidth]{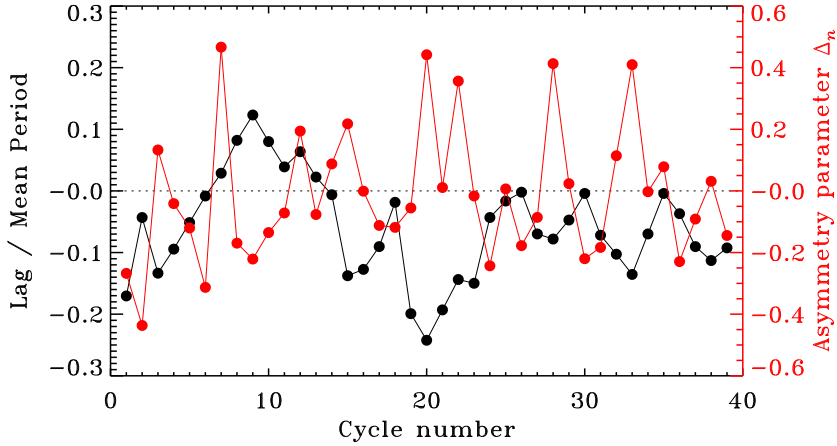}
\end{center}
\caption{Time sequence of the hemispheric asymmetry parameter $\Delta_n$ (in red), as defined through eq.~(\ref{eq:asym}) for the toroidal proxy, calculated for the 39 cycles in the simulation.
The corresponding time sequence of hemispheric lag, defined as the difference between the time of cycle minimum in the north minus the time of cycle minimum in the south, is plotted in black.
\label{fig:Dario2}}
\end{figure}

Each pair of hemispheric proxy time series correlate rather well as a whole, reflecting primarily the good hemispheric synchrony of polarity reversals. However, the peak amplitudes for individual cycles in the north and south do not
($r=-0.24$ and $+0.23$ for the toroidal and dipolar proxies, respectively). Within an hemisphere, the peak amplitudes of the toroidal and dipolar proxies do correlate well with one another ($r=+0.58$ and $+0.66$ in the north and south, respectively). This is suggesting that once underway, cycles in each hemisphere rise and saturate independently of one another. We define as follows an asymmetry parameter $\Delta_n$ for cycle $n$ in terms of the peak amplitudes $A^N_n,A^S_n$ of the proxies in the northern and southern hemispheres. This parameter if similar to the normalized asymmetry parameter presented in \S 2.1:
\begin{eqnarray}
\label{eq:asym}
\Delta_n={A^N_n-A^S_n\over A^N_n+A^S_n}~,\qquad n=1,...,39
\end{eqnarray}
so that $\Delta_n\to +1$ ($-1$) if the northern (southern) hemisphere dominates, and $\Delta_n=0$ if both hemispheres exhibit the same cycle amplitude. The time sequence of $\Delta_n$ values for their toroidal proxies is plotted in red on Figure \ref{fig:Dario2}; the corresponding sequence for the polar field proxy is very much similar. The asymmetry parameter reaches values as large as $\pm 0.5$ for some cycles, which is much larger than for the sunspot cycles on Fig.~\ref{fig:mcint2}.

Over the 20 full magnetic cycles covered in the 1600$\,$yr of this specific simulation, The average hemispheric (half-)cycle periods are virtually identical, $40.65\pm 1.81\,$yr in the north, and $40.57\pm 1.65\,$yr in the south. Moreover, as already noted,
polarity reversals maintain a very good synchrony in both hemispheres, hinting at some
significant level of cross-hemispheric coupling, even though the hemispheric cycle amplitudes are not well-correlated. The time sequence in black on Figure \ref{fig:Dario2} shows the
time sequence of hemispheric lag values for successive cycles defined here as the time of cycle minimum in the north minus that in the south, as defined by the toroidal proxies
(so that a negative lag indicate that the north leads the south). The average lag is 2.3$\,$yr over the whole simulation, or about 6\% of the cycle period. The final cumulative lag after 39 cycles amounts to a mere 10\% of the cycle period for the toroidal proxy, and only 2\% for the polar field proxy (not shown). At first glance the lag seems to execute a form of random walk, with
the lag increasing or decreasing by the same amount (within a factor of two)
from one cycle to the next. The asymmetry parameter, in contrast, shows
a much more abrupt cycle-to-cycle variation.  The plot is also suggestive of an anticorrelation
of the amplitude asymmetry parameter, $\Delta_n$, with the lag values, but the correlation coefficient turns out to be rather low, $r=-0.31$.

\section{Coupling Mechanisms\label{sec:coupling}}

The various models and simulations reviewed in the preceding section exhibit varying levels of cross-hemispheric coupling, in part because different physical coupling mechanisms are at play, and/or operate in distinct physical regimes. In this section we focus specifically on these mechanisms, and discuss under which circumstances they can (or cannot) achieve cross-hemispheric coupling.

\subsection{Magnetic diffusivity}

The last term on the RHS of eq.~(\ref{eq:mfmhd}) embodies resistive dissipation of the large-scale magnetic field. In a situation where $\beta\gg\eta$ and is dependent on position, this resistive term can be developed as
\begin{equation}
\label{eq:mfmhddif}
-\nabla\times( \beta\nabla\times\avB)=
-(\nabla\beta)\times(\nabla\times\avB) +\beta\nabla^2 \avB~,
\end{equation}
The first term on the RHS is known as diamagnetic transport; it is fundamentally distinct from the turbulent pumping introduced in \S\ref{sec:dynamo}, as it can arise in a situation were turbulence is isotropic (the $\bfalpha$ and $\bfbeta$ tensors are diagonal), provided the $\beta$ coefficient varies with position. True turbulent pumping, in contrast, materializes only when turbulence is anisotropic, as captured in the off-diagonal component of the $\bfalpha$-tensor. The second term on the RHS has the form of a classical Fickian (linear) diffusion, with $\beta$ acting as a diffusion coefficient. This offers a prime mechanism to achieve cross-hemispheric
coupling. However, of all dynamo ingredients required in mean-field and mean-field-like
dynamo models, the turbulent magnetic diffusivity $\beta$ is perhaps the most difficult quantity to estimate reliably from first principles, and currently the one for which the least direct
observational constraints are available.

While most can agree that it is turbulent and anisotropic, i.e. there is greater meridional diffusion than radial diffusion due to the rotational influence on convection, we can agree on little else.  For example, measurements of $\beta$ at the surface determined from observations vary depending on the feature and size-scale being studied.   The values for magnetic diffusion reported from observational studies are:  600 $km^{2}s^{-1}$ for supergranules \citep{Simon1995}, to 200 $km^{2}s^{-1}$ for small-scale magnetic elements \citep{Komm1995}, 60 $km^{2}s^{-1}$ for granular flows \citep{Berger1998}, and
1-5 $km^{2}s^{-1}$ for high-resolution plage flows \citep{Chae2008}.  The disagreement of the correct range of magnetic diffusion as a function of radius is equally apparent in modeling and simulation efforts.  Two examples of assumed diffusion values (in units of $cm^{2} s^{-1}$) as a function of solar radius are $6\times 10^{11}$ near the surface, peaking at 0.93 R$_\odot$ with a value of $14\times 10^{11}$, declining gradually to $1\times 10^{11}$ at $0.72 R_\odot$ \citep{Pipin2013}.  In contrast, \citet{Dikpati2004} assume a diffusivity profile that peaks at the surface at $2\times 10^{12}$ (a supergranular value) declining to a constant value of $5\times 10^{10}$ from $0.73-0.90R_\odot$.  A $\beta$ of $2\times 10^{11} cm^{2}s^{-1}$ means the magnetic field diffuses the depth of the convection zone in 50 years, but diffuses to mid-depth in 10 years.

Consider, in the descending phase of the solar cycle, the two toroidal flux systems located on either side of the equatorial plane, peaking at $\pm 8^\circ\,$latitude (say). This corresponds to a linear distance $L \simeq 0.07\,$R$_\odot$. The timescale for these two structures to diffusively
annihilate is $\tau=L^2/\beta$; this is equal to the solar cycle period $\simeq 10\,$yr
for $\beta\simeq 10^{13}\,$cm$^2\,$s$^{-1}$. For a turbulent diffusivity in
excess of this value, one can thus expect diffusive cross-hemispheric coupling
to act on timescales shorter than the cycle period, i.e., strong coupling. Conversely,
for values of $\beta$ much smaller, the diffusive cross-hemispheric coupling
is correspondingly weaker. In this latter situation, any hemispheric
lag or amplitude asymmetry, of whatever origin, can endure for many magnetic cycles
(see, e.g., Dikpati \& Gilman 2004; Charbonneau 2007; Chatterjee, Nandy and Choudhuri, 2004). These two physical regimes have been dubbed ``diffusion-dominated'' and ``advection-dominated'' (see, e.g., \cite{Yeates2008}). Transequatorial diffusive transport also has a strong impact on the parity of the dynamo modes, with high magnetic diffusivity favoring the odd (dipolar) mode, characterized by the longest spatial scale, while in the advection-dominated
regime the even (quadrupolar) parity is imprinted on the dynamo \citep{Hotta2010}.

The distinction between these two physical regimes becomes particularly important in the so-called flux-transport dynamos (see Karak et al. review in this volume), in which the meridional flow drives the equatorward drift of the deep-seated toroidal magnetic field ultimately giving rise to sunspots (at least according to the prevalent view on the matter; see Schmieder et al. review on flux emergence, this volume). Even then high diffusivity is no universal panacea.  For example, \cite{Chatterjee2006} have studied dynamo solutions using a Babcock-Leighton-type
flux transport model in which small, temporally steady differences are introduced
between hemispheres in the amplitude of the meridional flow (1\% hemispheric
difference) and/or in the magnitude of the surface source term (2\% hemispheric
difference). Even for such small imposed hemispheric differences, persistent phase lag
and even distinct hemispheric periodicities can develop. Typically, in their model higher diffusivity is found to help couple the solutions in both hemispheres and establish a single period. These authors also point out that the presence of a weak turbulent $\alpha$-effect operating throughout though the convection zone also helps to keep the solution
locked in a dipolar parity, in agreement with earlier findings (Dikpati \& Gilman 2001).

Intermittency in the advection-dominated regime also tends to show strong hemispheric
asymmetries, with one or the other hemisphere entering Grand Minima state while
the other remains in ``normal'' cyclic mode (see, Charbonneau 2007 and Passos et al.~2014
for specific examples). Once again diffusive cross-hemispheric coupling can alleviate in
part this difficulty, and synchronize grand minima episodes in both hemispheres.
This behavior was explored in \cite{Charbonneau2005}, using a simple but well-validated
one-dimensional iterative map describing the variation of successive cycle amplitudes. More elaborate mean-field and mean-field-like models using higher magnetic diffusivity values, i.e. operating closer to or in the diffusion-dominated regime, tend to produce better hemispheric synchrony in Grand Minima. They also show a tendency for recovery to start preferentially
in one hemisphere (see Fig 8 in Passos et al.~2014), as was observed at the end
of the Maunder Minimum (Ribes \& Nesmes-Ribes 1993). This latter behavior
had been invoked in support of the ideas that deterministically-driven
parity modulation lead to Grand Minima (Sokoloff \& Nesmes-Ribes 1994;
Beer et al.~1998). Here it materializes simply because as one hemisphere shuts down, diffusive leakage from the other can shut it down as well after some temporal delay depending on the magnitude of turbulent diffusivity, and likewise following recovery in one hemisphere.

\subsection{Transequatorial convective flows and Poynting flux}

In mean-field and mean-field-like dynamo models, transequatorial transport of magnetic field by turbulent convection operates exclusively via the (enhanced) turbulent dissipation term $\propto\beta$ in eq.~(\ref{eq:mfmhdbis}). Sustained cross-equatorial meridional flows
are ruled out by dynamical and symmetry considerations. Latitudinal turbulent pumping vanishes at the equator by symmetry, but can lead to the concentration of magnetic fields at low latitudes,
with diffusion then taking over for cross-equatorial exchange.

\begin{figure}
\begin{center}
\includegraphics[width=0.6\textwidth]{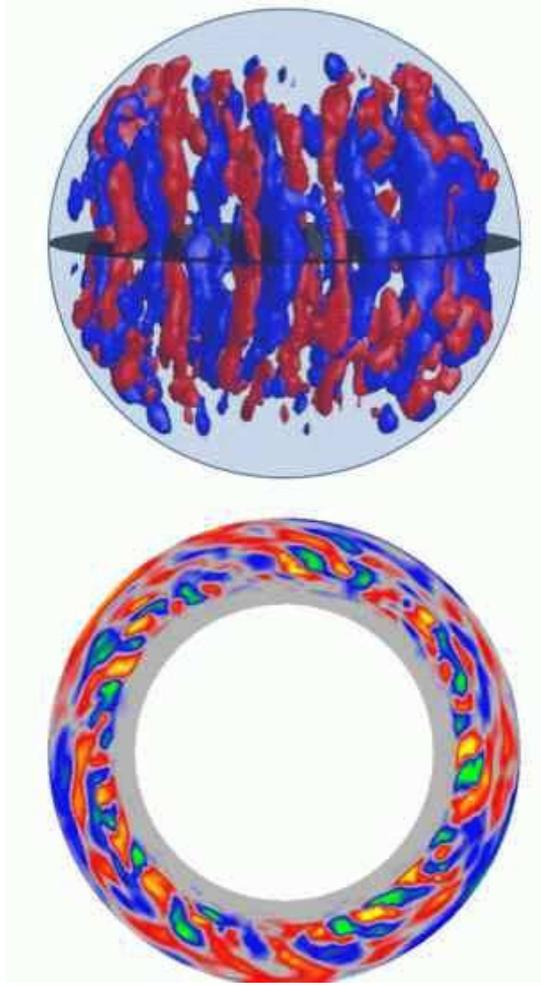}
\end{center}
\caption{Top: Snapshot of isosurfaces for the components of the vorticity parallel to the
rotation axis (red is positive, blue negative) in the EULAG-MHD simulation described previously in \S\ref{ssec:eulag}. The viewpoint is from slightly above the equatorial plane, shown as a dark disk. Bottom: latitudinal flow in the equatorial plane at one specific time in the same simulation (red-yellow: northward, blue-green southward, going from -$20$ to $+20$ m s$^{-1}$). The view is from the N-pole, down along the rotation axis.  This latitudinal flow crossing the equatorial plane ($\sim$1 m s$^{-1}$) is of
a magnitude similar to the roll speed of the cells in the equatorial plane, $\sim$10 m s$^{-1}$.
\label{fig:Paul1}}
\end{figure}

In global MHD simulations of rotating stratified convection, the situation turns out to be more complex, particularly in physical parameter regimes where rotation alters convection significantly. Outside of a cylinder tangent to the base of the convection zone, where no inner
boundaries can break the Taylor-Proudman constraint, convection organizes itself as a longitudinally-stacked series of latitudinally-elongated convective rolls (see Fig.~\ref{fig:Paul1}A), with the roll axes parallel to the rotation axis and sense of roll alternating from one to the next. These ``banana cells'' extend across the equator,
and drive an internal flow parallel to their axis alternating in their north-south direction from one roll to the next in longitude (viz.~Fig.~\ref{fig:Paul1}B; see also Busse 2002; Miesch \& Toomre 2009).

\begin{figure}
\includegraphics[width=1.0\textwidth]{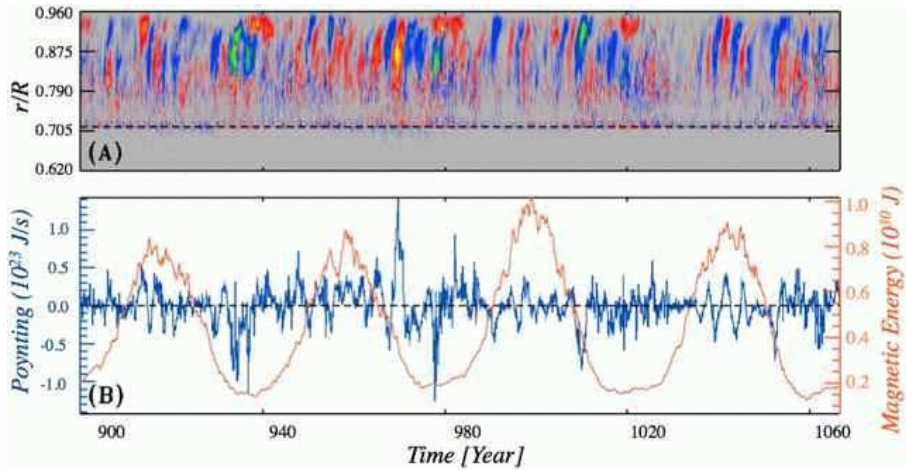}
\caption{Top: radius-time diagram of the zonally-averaged latitudinal component of the
Poynting flux in the equatorial plane, for the same EULAG-MHD simulation as in previous Figures. Positive values (in red-yellow) indicates a flux of electromagnetic energy from the southern into the northern hemisphere, and conversely for negative (in blue-green).
Bottom: time series of the latitudinal Poynting flux integrated over the full equatorial plane
(blue line), and of magnetic energy of the zonally-averaged toroidal magnetic field
(orange line), the latter indicative of
the phase of the large-scale magnetic cycle.
\label{fig:Paul2}}
\end{figure}

Although this trans-equatorial flow does not lead to any net mass exchange between
hemispheres, it can generate a net trans-equatorial Poynting flux (${\bf S}=\mu_0^{-1}{\bf E\times B}$) contributing to cross-hemispheric coupling, as illustrated on Figure \ref{fig:Paul2}. The top panel shows the zonally-averaged latitudinal component of the Poynting flux in the equatorial plane, in the form of a radius-time diagram spanning here 5 activity cycles. The bottom panel shows a time series of the net latitudinal Poynting flux across the equatorial plane (i.e., the top diagram integrated over depth), together with a time series of magnetic energy associated with the large-scale field $\avB$ (orange line).
The transequatorial Poynting flux is spatiotemporally very intermittent, and shows some clear short term quasi periodicities. Careful intercomparison of the top and bottom panels reveals that transequatorial activity is more pronounced in the descending phase of the large-scale magnetic cycle, and less so in the rising
phase. Integrating the Poynting flux over all depths (solid line on bottom panel) yield a very intermittent signal with zero mean, without much obvious imprint of the 40$\,$yr cycle period.

\begin{figure}
\begin{center}
\includegraphics[width=0.75\textwidth]{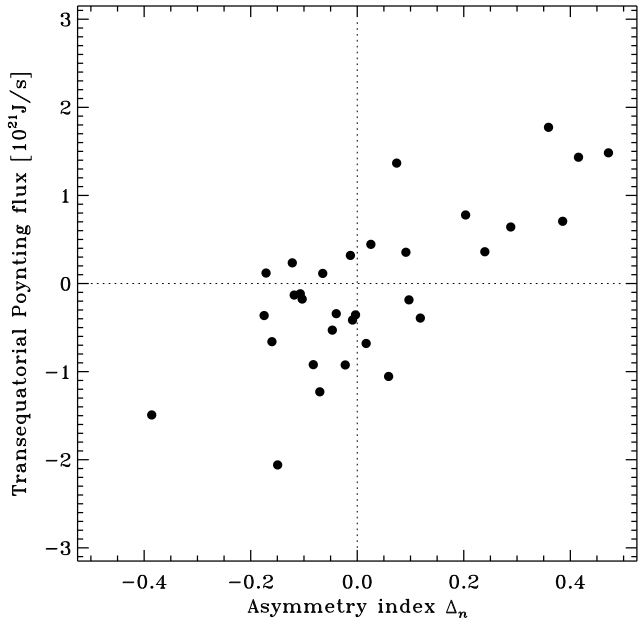}
\end{center}
\caption{Correlation plot of the trans-equatorial Poynting plot integrated over the equatorial plane in the descending phase of cycles, versus the asymmetry of hemispheric cycle amplitude, as defined in eq.~(\ref{eq:asym}). On average (linear correlation coefficient $r=+0.74$),
the Poynting flux carries electromagnetic energy from the weaker hemisphere into the dominant hemisphere, contrary to the ``diffusive'' expectation (see text). Numerical data from the EULAG-MHD simulation of Fig.~\ref{fig:Paul2} (see also Passos \& Charbonneau 2014).
\label{fig:Paul3}}
\end{figure}

The overall fluctuation level of the transequatorial Poynting flux on Fig.~\ref{fig:Paul2} does show some cycle-to-cycle variations. One might expect that these variations can be traced to the level of hemispheric asymmetry characterizing the amplitude of the large-scale toroidal magnetic field building up in each hemisphere for this simulation (viz.~Fig.~\ref{fig:Dario1} herein). More specifically, if the magnetic field is passively ``mixed'' across hemispheres by transequatorial flows, one would expect a net Poynting flux from the dominant hemisphere toward the weaker hemisphere. Figure \ref{fig:Paul3} shows the
transequatorial Poynting flux integrated over a 10$\,$yr-wide temporal window covering
the late descending phase of each cycle, plotted against the corresponding hemispheric asymmetry parameter $\Delta_n$ defined earlier (viz.~eq.~\ref{eq:asym}).
As in Fig.~\ref{fig:Paul2}, a positive Poynting flux indicates electromagnetic energy flux from the southern into the northern hemisphere, and a positive asymmetry parameter indicates dominance of the northern hemisphere toroidal magnetic field. The positive correlation ($r=+0.74$) runs contrary to the expectation that the stronger magnetic field of the dominant hemisphere leaks into the weaker hemisphere, which should translate here as an {\it anti}correlation. The observed correlation must instead materialize because the stronger magnetic field in the dominant hemisphere affects the turbulent transport coefficients in a manner such as to reduce the Poynting flux out of that hemisphere.  Under this view, hemispheric coupling becomes a fully nonlinear magnetohydrodynamical phenomenon, with
magnetically-mediated alterations of convective patterns dominating over passive, linear diffusive-like coupling.

Taken at face value, the positive correlation observed
in Fig.~\ref{fig:Paul3} means that hemispheric asymmetries should be amplified by the transequatorial Poynting flux, which does not seem to be happening in this simulation, as strongly asymmetric cycles (e.g. cycle 28 in the top panel of  Fig.~\ref{fig:Dario1}) are seldom followed by similarly asymmetrical cycles. Knowing the total magnetic energy content $E^N,E^S$ associated with the large-scale magnetic field in each hemisphere, and the total transequatorial Poynting flux ${\bar S}$, one can estimate
the timescale $\tau_\Delta$ required to equilibrate a 50\% difference (say) in hemispheric energy content:
\begin{eqnarray}
\label{eq:tauPoynting}
\tau_\Delta={E^N+E^S\over 4{\bar S}}~.
\end{eqnarray}
For the EULAG-MHD simulation analyzed here, typical values are $E^N\simeq E^S\simeq 6\times 10^{31}\,$J and ${\bar S}\simeq 2\times 10^{21}\,$J s${}^{-1}$ for the most asymmetric
cycles on Fig.~\ref{fig:Paul3}, leading to $\tau_\Delta\simeq 50\,$yr, i.e., a little larger than the
average cycle period in this simulation. One can but conclude that the transequatorial Poynting flux remains significant over cycle-timescales, at least in this specific simulation.

Interestingly, if the transequatorial Poynting flux integration is carried out only in the stable fluid layers underlying the convection zone, the expected ``diffusive'' anticorrelation is recovered ($r=-0.57$), but the net Poynting flux is nearly two orders of magnitude lower than that crossing the equatorial plane within the convection zone.
This is now more in line with purely diffusive behavior, reflecting the absence of
rotationally-aligned banana cells in the stable layer, and small-scale motions being much weaker therein than they are in the overlying convection zone. However, the corresponding hemispheric coupling timescale is correspondingly longer, and likely irrelevant over the time span of this simulation.

\subsection{Transequatorial meridional flows}

In mean-field models, the
large-scale meridional flow, i.e. the axisymmetric
flow component contained in meridional $[r,\theta]$ planes, is usually
considered steady and made of a single flow cell per hemisphere. This flow
component can also be extracted from MHD numerical simulations, which usually
reveals a more complex pattern of multiple flow cells per hemispheres and large
deviations from the zonal average as a function of longitude and time. The meridional flow also
shows significant temporal evolution, although on timescales longer than convection,
so that it can be legitimately be considered a physically meaningful flow component.

\begin{figure}
\centering
\includegraphics[width=0.75\textwidth]{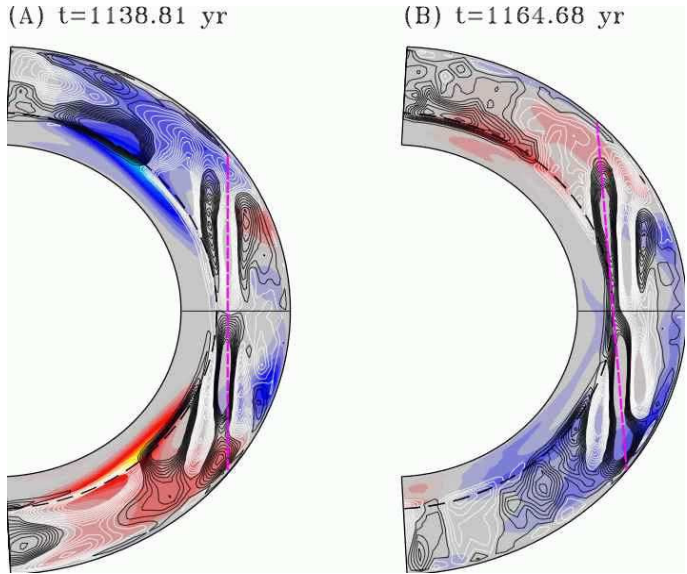}
\caption{The two panels show a snapshot of the meridional circulation stream function, $\psi$, taken at (A) the time of solar maximum (cycle 27 of Figure \ref{fig:Dario1}) and (B) the following minimum. The contour lines represent the complex cell structure of the meridional circulation where black indicates counter-clockwise rotation and white clockwise. These meridional cells are
overlaid on top of the zonal mean toroidal field (as on in panels (A) and (B) of figure \ref{fig:Dario0}). The dashed pink line is plotted to show deviations from vertical alignment.
\label{fig:cross-hemispheric_MC}}
\end{figure}

Following a procedure typically used in mean-field models, the zonally averaged components of the velocity  $u_r$ and $u_\theta$ can be used to build a stream function $\psi$ so that
${\bf v}_p = \frac{1}{\rho} (\nabla \times \psi \hat{e}_\phi)$, representing the cellular structure of the meridional flow, ${\bf v}_p$.
Figure \ref{fig:cross-hemispheric_MC} shows this stream function taken at the maximum of simulated solar cycle 27 (see figure \ref{fig:Dario1}A) and at the following minimum. In this simulation, at low latitudes the meridional flow shows a cell topology dominated by the imprint of  banana cells previously described (viz.~Fig.~\ref{fig:Paul1}). These flow cells tend to
extend symmetrically across the equator, as highlighted by the dashed purple vertical line in the figure. This vertical alignment tends to break around cycle minima with the cells merging near the equator and forming \textit{skewed} (not parallel to the rotation axis) trans-equatorial cells. The duration of the anti-symmetric coupling between cells is usually short,
of the order of 3 yr in the simulation, which is much less than the cycle period but much more
than the convective turnover time in the bulk of the convecting layers. Whether this is related to some
of the quasi-periodicities apparent on Fig.~\ref{fig:Paul2}B (viz. around $t=1040\,$yr), and the impact
it may have on cycle amplitude, is still under investigation.

Looking at the latest measurements of \cite{Zhao et al 2013}, the situation on the Sun seems to be different, mainly due to the presence of the large vertical rolls (see Fig.~\ref{fig:Paul1}, top panel) in the simulation affecting the development of solar-like meridional circulation.  Nevertheless, it is completely plausible that a similar mechanism might be operating in the Sun on a much weaker/smaller scale.

\subsection{Impulsive versus Steady-State Coupling}

Diffusive cross-hemispheric coupling, as embodied in the last term on the RHS of eq.~(\ref{eq:mfmhd}), is a continuous process operating gradually at a rate proportional to the local gradient of the large-scale magnetic field, with the turbulent diffusivity $\beta$ providing
the proportionality constant. Acting across the equatorial plane on the dipolar large-scale magnetic component generated in each hemisphere, diffusive coupling operates throughout the whole cycle, adjusting to whatever amplitude asymmetry or temporal lag may be developing across hemispheres.

The hemispheric coupling produced by convectively-driven transequatorial flows, in contrast, is anything but a slowly-varying, quasi-steady process. This quite evident upon cross-examination
of Figs.~\ref{fig:Paul2} and \ref{fig:Paul3} herein. The transequatorial Poynting flux averaged over the descending phase of the cycles is of order $10^{21}\,$J s$^{-1}$, yet the corresponding time series on Fig.~\ref{fig:Paul2}B reveals a high level of temporal intermittency, with the transequatorial flux often changing sign in the course of the same cycle and often peaking at values approaching $10^{22}\,$J s$^{-1}$ for time intervals of a few years. Moreover, Figure \ref{fig:Paul2}A shows that these surges are well-localized spatially, and typically take place in the upper half of the convecting layers in this simulation.

A recent paper by Cameron et al. (2014) sheds light on how effective the occasional active region that emerges close to the equator can be.  This is an impulsive, diffusive event which is contrary to the conventional picture retained of diffusion occurring slowly and steadily during the cycle.
Cameron et al. (2014) explores surface flux transport in the Babcock-Leighton model with an emphasis on inflows into the active region belts and active region emergence near the Equator.  They show that the flux transported by a few cross-equatorial flux plumes by tilted sunspot groups near the equator is important for the polar field amplitude and open flux at minima.  They argue that inflows to active regions (a nonlinear effect in the BL dynamo) and cross-equatorial flux plumes provides an explanation for the weakness of the polar fields at the end of solar cycle 23, which led to a weak cycle 24 (or more on these matters see review
by Hathaway et al., this volume).

\subsection{Toroidal Flux Cancellation in the Interior Across the Equator}

We propose that late in the solar cycle, the strong toroidal magnetic flux toroids at the bottom of the convection zone begin to interact across the Equator.  At this time, magnetic flux cancellation may actively couple the hemispheres  in a manner distinctly different than the mechanisms described in \S4.  This proposed coupling process, wherein the flux toroids in the interior actively cancel across the equator, has not been explored in detail in the literature.  It is our intention to follow-up this paper with future research including observations and modeling as to how this might proceed.

We limit ourselves to a brief depiction (with broad brush stokes) of the proposed process in the Sun.  Exactly when in the cycle this leading edge flux cancellation across the Equator would occur depends on how broad the toroidal bands (or magnetic wreaths) are in latitude.  Using sunspot location patterns and a forward model to match sunspot data from Cycles 21$-$23, Norton \& Gilman (2005) report that the toroidal bands are roughly 10$^{\circ}$ wide in latitude at the beginning of the solar cycle but become broader, possibly due to magnetic drag, to $\sim$20$^{\circ}$ wide late in the cycle.  If the bands in each hemisphere are 20$^{\circ}$ in latitude, then when the butterfly wing centroids are at 10$^{\circ}$ latitude, the leading edge of the butterfly wings begins to reconnect across the Equator.  The observation that sunspots avoid the equator has been attributed to the meridional circulation decreasing its equatorward flow and turning upwards toward the surface (Hathaway et al. 2003).  While the meridional flow is certainly seen at the surface to decrease to near\--zero values (errors due to $p$-angle changes preclude absolute value determination)  and increase dramatically as one moves away from the Equator, we argue that sunspots avoid the Equator more persistently after the toroidal bands meet at the Equator (deep under the surface). Meaning, there is a V-shaped hollow in the butterfly diagram early in the cycle before maxima, but there is also a smaller, less notable hollow that appears after cycle maxima, see the location of the red ovals in Fig. 1.  Simulations show that weak sunspots have a non-radial emergence flux trajectory through the convection zone i.e. they rise slightly pole-ward of the latitude in the interior that they were formed (Choudhury \& Gilman 1987).  This may keep sunspots away from the Equator.  However, some spots do emerge on or very near the Equator so it is worth considering other mechanisms that might explain it.

When the late$-$cycle void at the Equator appears in Fig. 1, the butterfly wings are eroding from the Equator upwards in latitude.   We propose that an active flux cancellation at depth may prevent sunspots from emerging near the Equator late in the cycle as well as assisting the coupling of the hemispheres.  This erosion at the Equator may explain why the latitudinal centroid of hemispheric sunspot locations is seen to move poleward (or retrograde) at times late in the cycle (see Figs. 2, 3 in Ternullo 2007, Fig. 4 in Norton \& Gilman 2004).   The centroid eventually moves equatorward again after a period of months or years.  A ``ghost thorax" of the butterfly diagram may indicate a time that the leading (or equatorward) edges of the toroidal bands cancel each other as opposing currents meet.  This flux cancellation would not look exactly like ohmic diffusion (too slow) or X-point magnetic reconnection (too fast) but could be an effective mechanism of hemispheric coupling.  When the hemispheres are significantly out of phase, the magnetic equator becomes offset in relation to the geometric equator (see Zolotova et al. 2010) in which case the bands meet up and cancel slightly above or below the geometric Equator.  More research is necessary to determine under what conditions flux cancellation at depth may occur and what the observational signatures would be, including whether the absence of sunspots near the Equator late in the cycle is significant.

\section{Discussion:  Observations versus Models}

To briefly point out the similarities and dissimilarities in hemispheric asymmetry in observations and the EULAG-MHD simulations, we provide the following list.

\begin{enumerate}
\item There is very good agreement in synchrony for polarity reversals as observed in the sunspot data and the simulations.
\item Low hemispheric phase lags are seen in both data and simulations.  Phases are less than two years or 20\% of cycle time in observations and less than 25\% of cycle period in simulations.
\item Both observations and models show a persistent phase lag for four cycles (in observations) and more (for simulations).
\item Both observations and modeling agree that magnetic flux crosses the equator during the mid to late declining phase of the cycle.
\item Simulations do not produce correct cycle length nor do they produce toroidal fields close enough to the Equator to create a solar-like butterfly diagram. It is conceivable that in this simulation, the banana cells and associated meridional
flow structures play a significant role in preventing the toroidal field from migrating toward the Equator, leading to deep-seated toroidal bands that effectively remain fixed
in latitude, in stark contrast to the solar cycle. This may, in turn, lead
to diminished hemispheric coupling and stronger amplitude asymmetries (next item).
\item Stronger amplitude asymmetries are found in the toroidal field in simulations, up to 40\% difference in the hemispheres in a single cycle. This is twice as asymmetric as the observed values which are maximally 20\% asymmetric.
\item In global MHD simulations, the dipole strength is usually
not a successful precursor for next cycle toroidal proxy, whereas it is known observationally
that the solar polar fields do have forecasting value
(Charbonneau \& Barlet, 2011, \cite{Munoz-Jaramillo2013b}). Because there is no surface flux transport in the simulation, the dipole source term as generated by active region decay with bipolar tilt angles is not incorporated.
Instead the dipolar fields that are generated are a direct consequence of the inner dynamo mechanism. This could have an impact on the degree of hemispheric coupling
characterizing the simulation.
\item  \cite{Munoz-Jaramillo2013b} showed that polar flux becomes a
better cycle predictor by taking advantage of the hemispheric polar field strength proxy.  Since the hemispheric polar field strengths are a significantly better precursor than the whole-Sun measure, this is one more indication that the hemispheric coupling is imperfect.
\end{enumerate}

\section{Conclusions}

A slow ($\sim$10 year timescale), whole-Sun magnetic diffusion rate alone may be too small to account for the strong coupling of the hemispheres as observed in the butterfly diagram.   Examination of the butterfly diagram also shows an absence of sunspots that emerge near the equator, but only after the toroidal bands have met up at the magnetic equator,  see the butterfly diagram in Fig. 1.  The circles indicate the void, or the ``ghost thorax" of the butterfly, where the wings have met at the equator, but then a sharp decrease in the number of low latitude spots being produced is evident.  We propose that the suppression of low latitude sunspots late in the solar cycle (the lack of sunspots near the equator) is an indication that the bands are undergoing an active flux cancellation in the interior that decreases the amount of flux available to produce sunspots. It is a type of forced or confined ohmic diffusion that is not part of our current conventional picture of hemispheric coupling.  The toroidal bands, at some point, meet across the equator and actively couple the hemispheres late in the solar cycle by cancelling out the flux in the opposite toroidal band.   When one uses the distribution of surface flux and a surface diffusion rate to calculate the cross-equatorial flux, the estimate is found to peak slightly after cycle maximum and then decrease (Cameron and Sch\"ussler, 2007; Norton and Gallagher, 2010) due to the avoidance of flux at the equator late in the cycle.  Again, we argue that cross-equatorial flux cancellation at depth late in the solar cycle is causing a dearth of magnetic flux at the surface at this time right around the Equator.

In the vast majority of extant mean-field dynamo models of the solar cycle, cross-hemispheric coupling is mediated by (linear) diffusive transport, i.e., the last term on the RHS of eq.~(\ref{eq:mfmhd}).
In models where this diffusive transport is very efficient, stochastically-driven amplitude asymmetries and hemispheric phase lags are rapidly erased unless the dynamo is operating in the mildly supercritical regime.
Consequently, significant hemispheric asymmetries can only be sustained through parity modulation of odd/even modes, excited either stochastically by turbulent convection
or deterministically through nonlinear magnetic backreaction on large-scale flows.

In contrast, in mean-field dynamo models operating in the weakly diffusive, advection-dominated regime, persistent hemispheric asymmetries and phase lags can be triggered and sustained by (relatively) weak, spatiotemporally uncorrelated stochastic forcing, even with a coherence time for these fluctuations much smaller than the cycle period.
In either cases, the paucity of observational constraints means that many critical parameters and/or functionals introduced in such models can be
freely adjusted so as to yield the desired dynamo behavior.
This does not diminish the value of such models as exploratory thinking tools,
but implies that their role is primarily descriptive, rather than predictive.

Global MHD simulations of solar convection producing large-scale magnetic cycles,
such as the EULAG-MHD ``millennium simulation''
used herein for illustrative purposes, are offering a new perspective on the problem.
Not only can they potentially replace observations to some extent in constraining
the parameter space of simpler forms of dynamo models, but they also open a window into the fully dynamical, multi-scale 3D regime which, by definition, is inaccessible to the mean-field formulation.
For example, it is through the careful analysis of MHD numerical simulations starting a little over a decade ago that turbulent pumping is resurfacing as a potentially key
ingredient for the spatiotemporal evolution of the solar large-scale magnetic field.
Another, discussed in this paper, is the importance for cross-hemispheric coupling of transequatorial flows associated with persistent convective structures, with it associated fully magnetohydrodynamical, non-diffusive transport of magnetic fields.

Arguably the most striking discrepancy between the hemispheric asymmetries characterizing the
EULAG-MHD simulation used in this paper or comparison to the solar cycle lies with
hemispheric cycle amplitudes, which show a N--S correlation much smaller
than observed. On the one hand, this might be understood upon noting that
the toroidal field bands in the simulation peak at mid-latitudes and show
very little equatorward propagation (see Fig.~\ref{fig:Dario0}); consequently,
these ``activity bands'' never meet at the equator, in contrast to what is seen in
the sunspot butterfly diagram (cf.~Fig.~\ref{fig:1}).
It is then perhaps natural to expect the real solar cycle to exhibit higher levels
of cross-hemispheric coupling. On the other hand, the sun's ability to sustain
over many cycles
a cross-hemispheric lag in the sunspot counts (viz.~Fig.~\ref{fig:zolo1})
puts an upper bound on the efficiency of whichever physical mechanism
is responsible for magnetic cross-hemispheric coupling.
Ongoing efforts to extend the hemispheric sunspot number time series
(Clette et al., this volume), as well as the sunspot butterfly diagram (Arlt et al., this volume)
all the way back to the beginning of the SSN record, would be most helpful
in providing tighter constraints to modelling and simulation efforts.

It is perhaps appropriate to close this review with a reality check. Despite staggering advances in computing power and algorithmic design, current global MHD simulations are still a long way from the solar parameter regime. Even the most solar-like large-scale magnetic cycles they produce still show important discrepancies with respect to the observed solar cycle. Moreover, many important processes such as sunspot emergence, with subsequent decay and surface
flux dispersal, still cannot be captured within such simulations due to the extreme disparity of the spatial and temporal scales involved. Nonetheless, these simulations are dynamically consistent on spatial and temporal scales that they do resolve, and as such their output can be treated as ``experimental data'' to explore the intricacies of dynamo action in thick, stratified, rotating shells of convectively turbulent electrically conducting fluid. This is the approach we have adopted here in the context of cross-hemispheric coupling; but more generally,
it represents a unique springboard towards broader investigations of dynamo action in the sun and stars.

\begin{acknowledgements}
We thank Nicolas Lawson for producing some Figures and the analysis leading to Fig.~\ref{fig:Paul3}.  We thank S.  McIntosh and N. Zolotova for allowing the reproduction of several figures regarding sunspot measures of hemispheric asymmetry.  We thank J. Janssens for maintaining and updating the Cycle 24 website. P. Charbonneau is supported by the Natural Sciences and Engineering Research Council of Canada. D. Passos acknowledges support from the Funda\c{c}\~{a}o para a Ci\^{e}ncia e Tecnologia (FCT) grant SFRH/BPD/68409/2010, CENTRA-IST and the University of the Algarve for providing office space.  A. Norton is supported by NASA Contract NAS5-02139 (HMI) to Stanford University. We thank the ISSI organizers for a most productive and insightful conference held in Bern in November 2013.
\end{acknowledgements}



\begin{thebibliography}{}
%
%

\bibitem[\protect\citeauthoryear{Antonucci et al}{1990}]{Antonucci 1990}
Antonucci, E., Hoeksema, J.T., Scherrer, P.H., 1990, ApJ, \textbf{360}, 296, (1990).

\bibitem[\protect\citeauthoryear{Arlt \etal}{2013}]{Arlt2013}
Arlt, R., Leussu, R., Giese, N., Mursula, K., Usoskin, I. G., MNRAS, \textbf{433}, 3165 (2013).

\bibitem[\protect\citeauthoryear{Balmaceda \etal}{2009}]{Balmaceda2009}
Balmaceda, L. A.,  Solanki, S. K., Krivova, N. A., Foster,  S.,
J. Geophys. Res. \textbf{114}, A07104 (2009),

\bibitem[\protect\citeauthoryear{Babcock}{1959}]{Babcock1959}
Babcock, H.D., ApJ, \textbf{130}, 364 (1959).

\bibitem[\protect\citeauthoryear{Beaudoin et al.}{2013}]{Beaudoin et al. 1993}
Beaudoin, P., Charbonneau, P., Racine, \'E., Smolarkiewicz, P.K.,
SolP, \textbf{282}, 335--360 (2013).

\bibitem[\protect\citeauthoryear{Becker}{1955}]{Becker 1955}
Becker, U., Z. Astrophys., \textbf{37}, 47 (1955).

\bibitem[\protect\citeauthoryear{Beer et al.}{1958}]{Beer et al. 1998}
Beer, J., Tobias, S.M., Weiss, N.O., Solar Phys, \textbf{181}, 237 (1998).

\bibitem[\protect\citeauthoryear{Benevolenskaya}{2007}]{Benevolenskaya2007}
Benevolenskaya, E.E., , Highlights of Astronomy, \textbf{14}, 273 (2007).

\bibitem[\protect\citeauthoryear{Berger \etal}{1998}]{Berger1998}
Berger, T.~E., L{\"o}fdahl, M.~G., Shine, R.~A., \& Title, A.~M., \apj, \textbf{506}, 439 (1998).

\bibitem[\protect\citeauthoryear{Boyer \& Levy}{1984}]{Boyer 1984}
Boyer, D.W., Levy, E.H., ApJ, \textbf{277}, 848, (1984).

\bibitem[\protect\citeauthoryear{Brandenburg and Subramanian}{2005}]{Brandenburg and Subramanian 2005}
Brandenburg, A., Subramanian, K., Phys.~Rep., \textbf{417}, 1--209 (2005).

\bibitem[\protect\citeauthoryear{Brown et al.}{2010}]{Brown et al. 2010}
Brown, B.P., Browning, M.K., Brun, A.S.,
Miesch, M.S., Toomre, J., ApJ, \textbf{711}, 424 (2010).

\bibitem[\protect\citeauthoryear{Brown et al.}{2011}]{Brown et al. 2011}
Brown, B.P., Miesch, M.S.,
Browning, M.K., Brun, A.S., Toomre, J., ApJ, \textbf{731}, id69 (2011).

\bibitem[\protect\citeauthoryear{Browning et al.}{2006}]{Browning et al. 2006}
Browning, M.K., Miesch, M.S.,
Brun, A.S., Toomre, J., ApJL, \textbf{648}, L157--160 (2006).

\bibitem[\protect\citeauthoryear{Bushby}{2003}]{Bushby2003}
Bushby, P., \mnras, \textbf{328}, 655 (2003).

\bibitem[\protect\citeauthoryear{Busse}{2002}]{Busse 2002}
Busse, F.H., Phys.~Fluids, \textbf{14}, (2002).

\bibitem[\protect\citeauthoryear{Ballester et al}{2005}]{Ballester at al 2005}
Ballester, J.L., Oliver, R., Carbonell, M., A\&A, \textbf{431}, 5L (2005).

\bibitem[\protect\citeauthoryear{Caligari \etal}{1995}]{Caligari1995}
Caligari, P., Moreno-Insertis, F., Sch\"{u}ssler, M. 1995,
\apj, 441, 886

\bibitem[\protect\citeauthoryear{Cameron et al}{2007}]{Cameron at al 2007}
Cameron, R.H., Sch\"ussler,  ApJ, \textbf{659}, 801 (2007).

\bibitem[\protect\citeauthoryear{Cameron et al}{2014}]{Cameron at al 2014}
Cameron, R.H., Jian, J., Sch\"ussler, M., Gizon, L., JGR, \textbf{119}, 680 (2014).

\bibitem[\protect\citeauthoryear{Carbonell et al}{1993}]{Carbonell at al 1993}
 Carbonell, M., Oliver, R., Ballester, J.L., A\&A, \textbf{274}, 497 (1993).

\bibitem[\protect\citeauthoryear{Carbonell et al}{2007}]{Carbonell at al 2007}
 Carbonell, M., Terradas, J., Oliver, R., Ballester, J.L., A\&A, \textbf{476}, 951 (2007).

\bibitem[\protect\citeauthoryear{Chae \etal}{2008}]{Chae2008}
Chae, J., Litvinenko, Y.~E., \& Sakurai, T., \apj, \textbf{683}, 1153 (2008).

\bibitem[\protect\citeauthoryear{Charbonneau}{2005}]{Charbonneau2005}
Charbonneau, P., \solphys , \textbf{229}, 345 (2005).

\bibitem[\protect\citeauthoryear{Charbonneau}{2007}]{Charbonneau2007}
Charbonneau, P., Adv. Sp. Res. , \textbf{40}, 899 (2007).

\bibitem[\protect\citeauthoryear{Charbonneau}{2010}]{Charbonneau 2010}
Charbonneau, P., LRSP, \textbf{7}, lrsp-2010-3 (2010).

\bibitem[\protect\citeauthoryear{Charbonneau}{2011}]{Charbonneau 2011}
Charbonneau, P.,  Barlet, G., J. Atmos.\& Sol. Terrestrial Phys.,\textbf{7}, \textbf{73}, 198 (2011).

\bibitem[\protect\citeauthoryear{Charbonneau}{2014}]{Charbonneau 2014}
Charbonneau, P., ARA\&A, \textbf{7}, \textbf{52}, in press (2014).

\bibitem[\protect\citeauthoryear{Chatterjee \etal}{2004}]{Chatterjee2004}
Chatterjee, P., Nandy, D., Choudhuri, A. R.,
\aap, \textbf{427}, 1019 (2004)

\bibitem[\protect\citeauthoryear{Chatterjee and Choudhuri}{2006}]{Chatterjee2006}
Chatterjee, P., Choudhuri, A. R., \solphys, \textbf{239}, 29 (2006).

\bibitem[\protect\citeauthoryear{Choudhuri}{1987}]{Choudhuri1987}
Choudhuri, A.R., Gilman, P.A., \apj, \textbf{316}, 788 (1987).

\bibitem[\protect\citeauthoryear{Choudhuri}{1992}]{Choudhuri1992}
Choudhuri, A.R., \aap, \textbf{253}, 277 (1992).

\bibitem[\protect\citeauthoryear{Cossette et al.}{2013}]{Cossette et al. 2013}
Cossette, J.-F., Charbonneau, P., Smolarkiewicz, P.K.,
ApJL, \textbf{777}, idL29 (2013).

\bibitem[\protect\citeauthoryear{Dasi-Espuig \etal}{2010}]{DasiEspuig2010}
Dasi-Espuig, M., Solanki, S. K., Krivova, N., Cameron, R., Pe\~{n}uela, T.,
\aap, \textbf{518}, 10 (2010).

\bibitem[\protect\citeauthoryear{Dikpati and Gilman}{2001}]{DikpatiGilman2001}
Dikpati, M., Gilman, P. A.,
ApJ, \textbf{559}, 428--442 (2001)


\bibitem[\protect\citeauthoryear{Dikpati \etal}{2004}]{Dikpati2004}
Dikpati, M., de Toma, G., Gilman, P.~A., Arge, C.~N., \& White, O.~R., \apj, \textbf{601}, 1136 (2004).

\bibitem[\protect\citeauthoryear{Dikpati \etal}{2006}]{Dikpati2006}
Dikpati, M., de Toma, G., Gilman, P. A.,
Geoph. Res. Lett., \textbf{33}, L05102 (2006)

\bibitem[\protect\citeauthoryear{Durrant}{2003}]{Durrant2003}
Durrant, C.J., Wilson, P.R., Sol. Phys., \textbf{214}, 23 (2003).

\bibitem[\protect\citeauthoryear{Duvall et al}{1993}]{Duvall et al 1993}
Duvall, T. L., Jr., Jefferies, S. M., Harvey, J. W., and Pomerantz, M. A., Nature, \textbf{362}, 430 (1993).

\bibitem[\protect\citeauthoryear{Fan and Fang}{2014}]{FanFang2014}
Fan, Y., Fang, F., ApJ, \textbf{789}, 35 (2014).

\bibitem[\protect\citeauthoryear{Ghizaru et al.}{2010}]{Ghizaru et al. 2010}
Ghizaru, M., Charbonneau, P.,
Smolarkiewicz, P.K., ApJL, \textbf{715}, L133--137 (2010).

\bibitem[\protect\citeauthoryear{Goel and Choudhuri}{2009}]{Goel2009}
Goel, A., Choudhuri, A. R., Res. Astron. Astroph., \textbf{9}, 115 (2009).

\bibitem[\protect\citeauthoryear{Hathaway}{2003}]{hathaway1989}
Hathaway, D.H., Nandy, D., Wilson, R.M., and Reichmann, E.J., \apj, \textbf{589}, 665 (2003).

\bibitem[\protect\citeauthoryear{Hill}{1989}]{Hill 1989}
Hill, F., \apj, \textbf{343}, L69 (1989).

\bibitem[\protect\citeauthoryear{Hotta and Yokoyama}{2010}]{Hotta2010}
Hotta, H., Yokoyama, T., \apjl, \textbf{714}, L308 (2010).

\bibitem[\protect\citeauthoryear{Howe et al}{2013}]{Howe et al 2013}
Howe, R., Baker, D., Harra, L., van Driel-Gesztelyi, L., Komm, R., Hill, F. and Gonz{\'a}lez Hern{\'a}ndez, I., ASP Series, \textbf{478}, 291 (2013).

\bibitem[\protect\citeauthoryear{Hoyng \etal}{1994}]{Hoyng1994}
Hoyng, P., Schmitt, D., Teuben, L. J. W., \aap, \textbf{289}, 265 (1994).

\bibitem[\protect\citeauthoryear{Joshi and Joshi}{2004}]{Joshi2004}
Joshi, B., Joshi, A., Sol. Phys., \textbf{219}, 343 (2004).

\bibitem[\protect\citeauthoryear{Kapyla et al.}{2012}]{Kapyla et al. 2012}
K\"apyl\"a, P.J., Mantere, M.J., Brandenburg A.,
ApJL, \textbf{755}, idL22 (2012).

\bibitem[\protect\citeauthoryear{Kapyla et al.}{2013}]{Kapyla et al. 2013}
K\"apyl\"a, P.J., Mantere, M.J., Cole E., Warnecke J., Brandenburg A.,
ApJ, \textbf{778}, id41 (2013).

\bibitem[\protect\citeauthoryear{}{1980}]{}
Krause, F, R\"adler, K-H,
{\it Mean-field magnetohydrodynamics and dynamo theory}.
Oxford: Pergamon Press. 271 pp. (1980).

\bibitem[\protect\citeauthoryear{Komm \etal}{1995}]{Komm1995}
Komm, R.~W., Howard, R.~F., \& Harvey, J.~W., \solphys, \textbf{158}, 213 (1995).

\bibitem[\protect\citeauthoryear{Komm et al}{2013}]{Komm et al 2013}
Komm, R.,  Howe, R., Gonz{\'a}lez Hern{\'a}ndez, I., Hill, F., Bogart, R.S. and Haber, D., ASP Series, \textbf{478}, 217 (2013).

\bibitem[\protect\citeauthoryear{Komm et al}{2011}]{Komm et al 2011}
Komm, R.,  Howe, Hill, F., R., Gonz{\'a}lez Hern{\'a}ndez, I., Hill, F., and Haber, D., IOP Journal of Physics, \textbf{271}, 012077 (2011).

\bibitem[\protect\citeauthoryear{Li etal}{2002}]{Li et al 2002}
Li, K.J., Liu, X.H., Yun, H.S., Xiong, S.Y., Liang, H.F., Zhao, H.Z., Zhan, L.S., Gu, X.M.: 2002,
Publ. Astron. Soc. Pac., \textbf{54}, 629 (2002).

\bibitem[\protect\citeauthoryear{Longcope and Fisher}{1996}]{Longcope1996}
Longcope, D. W., Fisher, G. H., \apj, \textbf{458}, 380, (1996).

\bibitem[\protect\citeauthoryear{Marwan et al}{2007}]{Marwan2007}
Marwan, N., Romano, M.C., Thiel, M., Kurths, J., Phys. Rep., \textbf{438}, 237 (2007)

\bibitem[\protect\citeauthoryear{Maunder}{1904}]{mau04}
Maunder, E.W., \mnras, \textbf{64}, 747 (1904).

\bibitem[\protect\citeauthoryear{McClintock and Norton}{2013}]{McClintockNorton2013}
McClintock, B.H., and Norton, A.A., Solar Phys., \textbf{287}, 215, (2013).

\bibitem[\protect\citeauthoryear{McIntosh et al}{2013}]{McIntosh et al 2013}
McIntosh, S.W. et al., ApJ, \textbf{765}, 146 (2013).

\bibitem[\protect\citeauthoryear{Miesch and Toomre}{2009}]{Miesch and Toomre 2009}
Miesch, M.S.,and Toomre, J., ARFM, \textbf{41}, 317--340 (2009).

\bibitem[\protect\citeauthoryear{Miesch \etal}{2000}]{Miesch2000}
Miesch, M., Elliott, J. R., Toomre, J., Clune, T. L., Glatzmaier, G. A.,
Gilman, P. A., \apj, \textbf{532}, 593 (2000).

\bibitem[\protect\citeauthoryear{Mininni and G\'omez}{2002}]{Mininni2002}
Mininni, P., G\'omez, D. O., \apj, \textbf{573}, 454 (2002).

\bibitem[\protect\citeauthoryear{Moffatt 1978}{1978}]{Moffatt 1978}
Moffatt, H.K.,
{\it Magnetic field generation in electrically conducting fluids},
Cambridge: Cambridge University Press. 343 pp. (1978).

\bibitem[\protect\citeauthoryear{Moss et al}{1952}]{Mossetal1992}
Moss, D., Brandenburg, A., Tavakol, R., Tuominen, I., \aap, \textbf{265}, 843 (1992).

\bibitem[\protect\citeauthoryear{Mu\~{n}oz-Jaramillo \etal}{2012}]{Munoz-Jaramillo2012}
Mu\~{n}oz-Jaramillo, A., Sheeley, N. R., Zhang, J. Jr.,  Deluca, E. E.,
\apj, \textbf{753}, 146 (2012).

\bibitem[\protect\citeauthoryear{Mu\~{n}oz-Jaramillo \etal}{2013a}]{Munoz-Jaramillo2013a}
Mu\~{n}oz-Jaramillo, A., Dasi-Espuig, M., Balmaceda, L., DeLuca, E.,
\apjl, \textbf{767}, L25 (2013).

\bibitem[\protect\citeauthoryear{Mu\~{n}oz-Jaramillo \etal}{2013b}]{Munoz-Jaramillo2013b}
Mu\~{n}oz-Jaramillo, A., M., Balmaceda, L., DeLuca, E.,
\prl, \textbf{111}, 041106 (2013).

\bibitem[\protect\citeauthoryear{Murakozy and Ludmany}{2012}]{Murakozy and Ludmany 2012}
Murak\"{o}zy, J., and Ludm\'any, A., MNRAS, \textbf{419}, 3624 (2012).

\bibitem[\protect\citeauthoryear{Mursula and Hiltula}{2003}]{MursulaHiltula 2003}
Mursula, K., Hiltula, T., Geo. Rev. Letters, \textbf{30}, id2135 (2003).

\bibitem[\protect\citeauthoryear{Nelson et al.}{2013}]{Nelson et al. 2013}
Nelson, N.J., Brown, B.P., Brun, A.S., Miesch, M.S., Toomre, J., ApJ, \textbf{762}, id73 (2013).

\bibitem[\protect\citeauthoryear{Newton \& Milsom}{1955}]{Newton1955}
Newton, H.W., Milsom, A.S., MNRAS, \textbf{115}, 398 (1955).

\bibitem[\protect\citeauthoryear{Norton \& Gallagher}{2010}]{NortonGallagher2010}
Norton, A.A., and Gallagher, P.A., Solar Phys., \textbf{261}, 193 (2010).

\bibitem[\protect\citeauthoryear{Norton \& Gilman}{2005}]{NortonGilman2005}
Norton, A.A., and Gilman, P.A., ApJ, \textbf{630}, 1194 (2005).

\bibitem[\protect\citeauthoryear{Olemskoy \& Kitchatinov}{2013}]{Olemskoy2013}
Olemskoy, S. V., Kitchatinov, L. L., \apj, 777, 71 (2013).

\bibitem[\protect\citeauthoryear{Ossendrijver \etal}{1996}]{Ossendrijver1996}
Ossendrijver, A. J. H., Hoyng, P., Schmitt, D., \aap, \textbf{313}, 938 (1996).

\bibitem[\protect\citeauthoryear{Ossendrijver \etal}{2001}]{Ossendrijver2001}
Ossendrijver, A. J. H., Stix, M., Brandenburg, A., \aap, \textbf{376}, 713 (2001).

\bibitem[\protect\citeauthoryear{Ossendrijver}{2003}]{Ossendrijver 2003}
Ossendrijver, M.A.J.H.,
A\&ARev, \textbf{11}, 287-367 (2003).

\bibitem[\protect\citeauthoryear{Ossendrijver et al.}{2002}]{Ossendrijver et al. 2002}
Ossendrijver, M.A.J.H., Stix, M.,
Brandenburg, A., R\"udiger, G., A\&A, \textbf{394}, 735--745 (2002).

\bibitem[\protect\citeauthoryear{Passos \etal}{2014}]{Passos2014a}
Passos, D., Nandy, D., Hazra, S., Lopes, I., \aap, \textbf{562}, A18 (2014).

\bibitem[\protect\citeauthoryear{Passos \& Charbonneau}{2014}]{PassosCharb2014}
Passos, D., Charbonneau, P., \aap, \textbf{568}, A113, DOI 10.1051/0004-6361/201423700, (2014).

\bibitem[\protect\citeauthoryear{Pipin}{1999}]{Pipin1999}
Pipin, V. V., \aap, \textbf{346}, 295 (1999).

\bibitem[\protect\citeauthoryear{Pipin and Kosovichev}{2013}]{Pipin2013}
Pipin, V.~V., \& Kosovichev, A.~G., \apj, \textbf{776}, 36 (2013).

\bibitem[\protect\citeauthoryear{Pulkkinen}{1999}]{Pulkkinen1999}
Pulkkinen, P.J., Brooke, J., Pelt, J., Tuominen, I., A\&A, \textbf{341}, L43 (1999).

\bibitem[\protect\citeauthoryear{Racine \etal}{2011}]{Racine2011}
Racine, E., Charbonneau, P., Ghizaru, M., Bouchat, A., Smolarkiewicz, P. K.,
\apj, \textbf{735}, 46 (2011).

\bibitem[\protect\citeauthoryear{Rempel}{2006}]{Rempel 2006}
Rempel, M., Heliophysics, eds. CJ Schrijver \& GL Siscoe, 42--74 (2006).

\bibitem[\protect\citeauthoryear{Simon \etal}{1995}]{Simon1995}
Simon, G.~W., Title, A.~M., \& Weiss, N.~O., \apj, \textbf{442}, 886 (1995).

\bibitem[\protect\citeauthoryear{Smolarkiewicz and Charbonneau}{2013}]{Smolarkiewicz and Charbonneau 2013}
Smolarkiewicz, P.K., Charbonneau, P., JCP, \textbf{236}, 608--623 (2013).

\bibitem[\protect\citeauthoryear{Sokoloff and Nesme-Ribes}{1994}]{Sokoloff1994}
Sokoloff, D., Nesme-Ribes, E., \aap, \textbf{288}, 293 (1994).

\bibitem[\protect\citeauthoryear{Sp\"{o}rer}{1894}]{Sporer1894}
Sp\"{o}rer, G., ibid Nr., \textbf{32}, 10, 1 (1894).

\bibitem[\protect\citeauthoryear{Sun et al}{2011}]{Sun2011}
Sun, X., Liu, Y., Hoeksema, J.T., Kayashi, K.,  Zhao, X., Solar Phys., \textbf{270}, 9 (2011).

\bibitem[\protect\citeauthoryear{Temmer et al.}{2006}]{Temmer2006}
Temmer, M., Ryb\'ak, J., Bend\'ik, P., Veronig, A., Vogler, F., Otruba, W., P\"otzi, W., Hanslmeier, A., Astron. Astrophys. \textbf{447}, 735 (2006).

\bibitem[\protect\citeauthoryear{Ternullo}{2007}]{Ternullo2007}
Ternullo, M., Mem. S.A. It., \textbf{78}, 596 (2007).

\bibitem[\protect\citeauthoryear{Tobias}{1997}]{Tobias1997}
Tobias, S., \aap, \textbf{322}, 1007 (1997).

\bibitem[\protect\citeauthoryear{Tobias et al.}{2001}]{Tobias et al. 2001}
Tobias, S.M., Brummell, N.H., Clune, T.L., Toomre, J.,
ApJ, \textbf{549}, 1183--1203 (2001).

\bibitem[\protect\citeauthoryear{Virtanen and Mursula}{2014}]{Virtanen and Mursula2014}
Virtanen, I.I., and Mursula, K., ApJ, \textbf{781}, 99 (2014).

\bibitem[\protect\citeauthoryear{Waldmeierl}{1955}]{Waldmeier1955}
Waldmeier, M., "Ergebnisse und Probleme der Sonnenforschung", Leipzig, Geest, \& Portig, 2$^{nd}$ edition, (1955).

\bibitem[\protect\citeauthoryear{Yeates \etal}{2008}]{Yeates2008}
Yeates, A., Nandy, D., Mackay, D. H., \apj, \textbf{673}, 544 (2008).

\bibitem[\protect\citeauthoryear{Zhao et al}{2013}]{Zhao et al 2013}
Zhao, J., Bogart, R.S., Kosovichev, A.G., Duvall, T.L., Jr., Hartlep, T., ApJ, \textbf{774}, 29 (2013).

\bibitem[\protect\citeauthoryear{Zolotova et al}{2009}]{Zolotova2009}
Zolotova, N.V., Ponyavin, D.I., Marwan, N., Kurths, J., A\&A, \textbf{503}, 197 (2009).

\bibitem[\protect\citeauthoryear{Zolotova et al}{2010}]{Zolotova et al 2010}
Zolotova, N.V., Ponyavin, D.I., Arlt, R., and Tuominen, I.., Astron. Machr., \textbf{331}, 765 (2010).


\end{thebibliography}
\end{document}